\documentclass[12pt]{article}
\usepackage{amssymb,amsmath,latexsym, mathtools}
\usepackage{graphics, color, graphicx}
\usepackage{epsfig,psfrag,float}
\usepackage{hyperref} 
\usepackage{cleveref}
\usepackage{soul}

\hypersetup{
    colorlinks=true,       
    linkcolor=purple,   
    citecolor=blue,        
    filecolor=magenta,      
    urlcolor=blue           
}

\newcommand{\psla}{\mbox{$\not{\! p}$}} 
\newcommand{\tpsla}{\mbox{$\not{\! \tilde{p}}$}} 
\newcommand{\kslash}{\mbox{$\not{\! k}$}}
\newcommand{\lslash}{\mbox{$\not{\! l}$}}

\newcommand{\pslash}{\mbox{$\not{\! p}$}}
\newcommand{\ppslash}{\mbox{$\not{\! {p^{\,\prime}}}$}}
\newcommand{\tkslash}{\mbox{$\not{\! \tilde{k}}$}}
\newcommand{\tpslash}{\mbox{$\not{\! \tilde{p}}$}}
\newcommand{\tlslash}{\mbox{$\not{\! \tilde{l}}$}}
\newcommand{\tppslash}{\mbox{$\not{\! \tilde{p}}^{\,\prime}$}}
\usepackage{float}
\usepackage{axodraw4j}
\usepackage{pstricks}
\usepackage{color}
\usepackage{soul}
\usepackage[normalem]{ulem}
\usepackage[title]{appendix}
\usepackage{caption}
\usepackage{subcaption}
\providecommand{\keywords}[1]
{
  \small	
  \textbf{{Keywords :}} #1
}
\usepackage{titlesec}
\usepackage{multirow}
\newcounter{daggerfootnote}
\newcommand*{\daggerfootnote}[1]{%
    \setcounter{daggerfootnote}{\value{footnote}}%
    \renewcommand*{\thefootnote}{\fnsymbol{footnote}}%
    \footnote[2]{#1}%
    \setcounter{footnote}{\value{daggerfootnote}}%
    \renewcommand*{\thefootnote}{\arabic{footnote}}%
    }

\newcounter{asteriskfootnote}
\newcommand*{\asteriskfootnote}[1]{%
    \setcounter{asteriskfootnote}{\value{footnote}}%
    \renewcommand*{\thefootnote}{\fnsymbol{footnote}}%
    \footnote[1]{#1}%
    \setcounter{footnote}{\value{asteriskfootnote}}%
    \renewcommand*{\thefootnote}{\arabic{footnote}}%
    }
\numberwithin{equation}{section} 
\usepackage{tabularx}
\begin{document}

\begin{center}
{\bf \large One loop $\beta$ functions for mirror noncommutative Euclidean  \\ Yukawa sector within electroweak-scale right handed neutrinos~model} \\
\vspace{0.4cm} {\large K. A. Bouteldja $^{1,2,}$\protect\asteriskfootnote{Contact author : bouteldja\_abderrahmane@univ-blida.dz, c.bouteldja@yahoo.fr}, N. Bouayed $^{1,}$\protect\daggerfootnote{Contact author : nn.bouayed@gmail.com}, S. Kouadik $^{3}$ and F-Z Ighezou $^{2}$}

\vspace{.3cm} ${}^1${\em \small Laboratoire de Physique THéorique et de l'Interaction Rayonnement Matière, Département de Physique, Faculté des Sciences, Université Saad Dahlab Blida 1, 09000 Blida,  Algeria} \\
\vspace{.3cm} ${}^2${\em \small Laboratoire de Physique Théorique et de didactique, Faculté de Physique, Université des Sciences et de la Technologie Houari Boumediene, 16111 Bab-Ezzouar, Algeria} \\
\vspace{.3cm} ${}^3${\em \small Laboratoire de Physique des Techniques Expérimentales et ses Applications, Faculté des Sciences, Université Yahia Fares de M\'ed\'ea, 26000 M\'ed\'ea, Algeria} \\
\end{center}

\renewcommand{\thefootnote}{\arabic{footnote}}
\setcounter{footnote}{0}

\begin{abstract}%
In this paper, we derive the renormalization scale dependence of noncommutative mirror Yukawa couplings.
To achieve this, we first formulate an Euclidean noncommutative version of the Yukawa sector within the electroweak-scale mirror right handed neutrinos model. Then, we calculate the noncommutative one loop $\beta$ functions of Yukawa couplings for mirror fermions involved in this model, by taking advantage of the Slavnov Taylor identities for the universal mirror Yukawa couplings, and by using the noncommutative vulcanised scalar and spinor propagators that prevent the UV/IR mixing. 
This leads us to a system of six cubic coupled first order differential equations that depend only on the mirror Yukawa couplings and not on the noncommutative deformation and vulcanised parameters. We solve this system numerically for different initial conditions to get the evolution of the mirror Yukawa couplings in terms of the renormalization scale. Furthermore, we discuss the link to the commutative case and analyze the occurrence of the Landau pole for some specific sets of initial conditions.
\end{abstract}

\section{Introduction} \label{introduction}
The standard model (SM) of particle physics is the current well experimentally established model that describes the electroweak and strong fundamental interactions and the mass generation of most fundamental particles \cite{ATLAS-Higgs-2022, CMS-Higgs-2022}. 
However, some observations such as the neutrino masses and mixing and the baryon asymmetry, and also some expected features like the anticipated unification of the SM with the gravitational theory at the Planck scale, can't be accommodated within the SM. This makes it widely believed that the SM is an effective theory valid till the TeV scale \cite{R-2019}. Beyond this energy scale, a more general theory should hold. This paves the way to a wide range of beyond standard model (BSM) extensions that extend the particle sector, the underlying dynamic, or the spacetime structure \cite{R-2019}. 

Among these extensions, the electroweak-scale mirror right handed neutrinos model (EW$\nu_R^M$) \cite{EWRH-Hung-2007} is of particular interest. This model built on the commutative Minkowski spacetime, preserves the same gauge group as in the SM but extends the scalar sector and adds mirror fermions. Moreover, it allows the non sterile right handed neutrinos to get mass in the electroweak energy scale~\cite{EWRH-Hung-2007}. This feature makes it important to deepen the investigation of the Yukawa coupling behavior of these right-handed neutrinos in terms of the energy scale.
To shed more light on the electroweak symmetry breaking mechanism involved in the mirror fermion sector within this commutative model, Le and Hung in~\cite{EWRH-Hung-2014} have calculated the one loop $\beta$ functions of the Yukawa coupling of the right handed neutrinos, as well as of the other mirror fermions embroiled in this model. This allowed them to introduce mirror fermion condensate states, at an energy scale close to that of the Landau pole presented by the Yukawa couplings of mirror fermions, which manifests itself on the TeV scale \cite{EWRH-Hung-2016}. 
Furthermore, it's important to note that this model fits well with the observed LHC Higgs particle~\cite{HHK-2015} and the observed LHCb muon transmutation \cite{Hung-2008, HLTY-2018}. This makes it a good candidate to be efficient at the energies explored by the LHC.

However, the prospect of a noncommutative version of the Yukawa mirror sector of this model becomes relevant in the context of future colliders \cite{FCC-2019,CEPC-2018-V1}. These colliders will probe the electroweak symmetry breaking mechanism more deeply, starting to explore the low sector in the intermediate range between the LHC scale and the Planck scale, where new physics correlated with the structure of spacetime should manifest itself.
Furthermore, a noncommutative version holds promise in shedding more light on challenging nonperturbative physical phenomena, such as the dynamical electroweak breaking symmetry. This adds to the fact that during the last two decades, looking for the renormalization group equations in the noncommutative spacetime era has gained importance. In particular, some noncommutative $\beta$ functions have been investigated especially for scalar QED in \cite{GBRN-2017}, fermionic QED and QCD in \cite{Ettefaghi_2010}, scalar Gurau Model in \cite{BenGeloun_2008}, Gross-Neveu model in\cite{Lakhoua_2007}, and $\phi^4$ theory in \cite{Grosse_2004}. Recently special attention has been paid to the noncommutative extension of the Yukawa interaction \cite{Gesteau_2020}.
Hence, to contribute to dealing with renormalization group equations in the noncommutative spacetime, we are especially interested in studying, to the leading radiative corrections, the effect of the noncommutative spacetime on the behavior of the EW$\nu_R^M$ mirror's Yukawa couplings as a function of the renormalization scale.

First, we construct the corresponding Euclidean Yukawa noncommutative Lagrangian, by replacing in the Euclidean version of the EW$\nu_R^M$, the usual fields product by the Moyal star product \cite{Moyal-product-2008, JMSSW-2001} that accounts properly for the noncommutative spacetime effect, and by taking into account all involved field permutations.

Moreover, it's well known that loop calculations in a noncommutative spacetime are confronted with the existence of a correlation between the ultraviolet (UV) and infrared (IR) regions of momentum space induced by the non-planar Feynman diagrams due to the non-local character inherent to the star product~\cite{MRS-1999, GLL-2000,S-2003}.
To tackle this mixing problem without spoiling translation invariance, Gurau et al. \cite{Gurau-2009} and subsequently, Bouchachia et al \cite{Bouchachia-2015} introduced non-local counter terms in the vein of the vulcanisation method~\cite{GW-2003,GW-2005,GW2-2005,FVT-phd}, which leads to a vulcanised Euclidean scalar and spinor propagators that prevent the UV/IR mixing in a $\phi^4$ theory supplemented by a Yukawa interaction, and consequently restores the renormalisability of the theory. 

Therefore, making use of these vulcanised propagators, we are well equipped to undertake the calculation of the noncommutative $\beta$ functions of the mirror Yukawa couplings to the one loop order within the proposed mirror Yukawa sector of the Euclidean noncommutative electroweak scale mirror right handed neutrinos model (ENC-EW$\nu_R^M$).

This paper is structured as follows. 
In the next section, we briefly review the EW$\nu_R^M$ model in Minkowski spacetime, then switch to Euclidean spacetime and formulate the noncommutative Euclidean Yukawa Lagrangian of the electroweak-scale right handed neutrinos model.
In Section \ref{section-3}, we express the mirror Yukawa Lagrangian in terms of renormalization constants and renormalized couplings and fields, then establish the Slavnov Taylor identities for mirror Yukawa couplings, and formulate the relation that connects the $\beta$ function to the counter terms involved in the mirror Yukawa interaction.
Section \ref{section-4} establishes the expression of the necessary one loop order ultraviolet counter terms, by applying the noncommutative vulcanised Feynman rules to the selected processes and calculating the corresponding radiative corrections, the computational details of which are reported in the Appendix.
In Section \ref{section-5}, we derive in terms of the mirror Yukawa couplings the corresponding expressions of the $\beta$ function. This leads us to the establishment of a system of six first order coupled nonlinear differential equations. Furthermore, the connection with the commutative case is discussed.
Section \ref{section-6} is devoted to the extraction and the discussion of the numerical solutions of the differential equation system. We draw the dependence of the mirror Yukawa couplings on the renormalization scale. We also report a link to the commutative results of Le-Hung~\cite{EWRH-Hung-2014, EWRH-Hung-2016} and discuss the position of the occurrence of the Landau pole. Finally, we draw our concluding~remarks.

\section{The model formulation: } \label{section-2}
\subsection{A brief review of the Minkowski commutative EW\texorpdfstring{$\nu_R^M$}{} model}
The EW$\nu_R^M$ is an extension of the Standard Model that preserves the gauge sector but extends the spinor sector by right handed mirror leptons $l_R^M$ and right handed mirror quark $q_R^M$ doublets, left handed mirror leptons $e_L^M$ and left handed mirror quarks $u_L^M$ and $d_L^M$ singlets. The scalar sector is also extended by a singlet $\varphi_{_S}$, a new doublet $\Phi_{2M}$, and two triplets $\chi$ and $\xi$ fields as depicted in Table \ref{ewrhn-table}.
\begin{table}[h!] 
\begin{center}
\scalebox{.75}{
\begin{tabular}{cccc}\hline
SM fermion fields & $SU(3)_c \otimes SU(2)_W \otimes U(1)_Y$ &  Mirror fermion fields &  $SU(3)_c \otimes SU(2)_W \otimes U(1)_Y$ \\ \hline \\
$l_{_L}$ =$ \begin{pmatrix}
\nu_{e_{_L}} \\ e_{_L}
\end{pmatrix}$ 
&$\begin{pmatrix}
1, 2, - \dfrac{1}{2}
\end{pmatrix}$ & $l_{\scriptscriptstyle R}^{\scriptscriptstyle M}$ =$ \begin{pmatrix}
\nu_{e_R}^{\scriptscriptstyle M} \\ e_{\scriptscriptstyle R}^{\scriptscriptstyle M}
\end{pmatrix}$ & $\begin{pmatrix}
1, 2, - \dfrac{1}{2}
\end{pmatrix}$
\\ \\
$q_{_L}$ =$ \begin{pmatrix}
u_{_L} \\ d_{_L}
\end{pmatrix}$ 
&$\begin{pmatrix}
3, 2, \dfrac{1}{6}
\end{pmatrix}$ & $q_{\scriptscriptstyle R}^{\scriptscriptstyle M}$ =$ \begin{pmatrix}
u_{\scriptscriptstyle R}^{\scriptscriptstyle M}\\ d_{\scriptscriptstyle R}^{\scriptscriptstyle M}
\end{pmatrix}$ & $\begin{pmatrix}
3, 2, \dfrac{1}{6}
\end{pmatrix}$
\\ \\
$e_{_R}$  & $\begin{pmatrix} 1, 1, -1 \end{pmatrix}$ 
& $e_{\scriptscriptstyle L}^{\scriptscriptstyle M}$  & $\begin{pmatrix} 1, 1, -1 \end{pmatrix}$
\\ \\
$u_{_R}$  & $\begin{pmatrix}3, 1, \dfrac{2}{3} \end{pmatrix}$ 
& $u_{\scriptscriptstyle L}^{\scriptscriptstyle M}$  & $\begin{pmatrix}3,  1, \dfrac{2}{3} \end{pmatrix}$
\\ \\
$d_{_R}$  & $\begin{pmatrix}3, 1, -\dfrac{1}{3} \end{pmatrix}$ 
& $d_{L}^{M}$  & $\begin{pmatrix}3, 1, -\dfrac{1}{3} \end{pmatrix}$
\\ \\ \hline
SM scalar fields & $SU(3)_c \otimes SU(2)_W \otimes U(1)_Y$ &  Extra-SM scalar fields &  $SU(3)_c \otimes SU(2)_W \otimes U(1)_Y$ \\ \hline \\
  &  
& $\varphi_{_S}$  & $\begin{pmatrix} 1,1, 0 \end{pmatrix}$
\\ \\
$\Phi_2 = \begin{pmatrix}
\phi_2^+ \\ \phi_2^0
\end{pmatrix} $ & $\begin{pmatrix} 1,2, \dfrac{1}{2} \end{pmatrix}$ 
& $\Phi_{2M} =\begin{pmatrix}
\phi_{2M}^+ \\ \phi_{2M}^0
\end{pmatrix}$  & $\begin{pmatrix} 1,2, \dfrac{1}{2} \end{pmatrix}$
\\ \\
 & 
& $\tilde{\chi} = \dfrac{1}{\sqrt{2}} \, \Vec{\tau}.\Vec{\chi} =\begin{pmatrix}
\frac{1}{\sqrt{2}} \chi^+ & \chi^{++} \\
\chi^0 & -\frac{1}{\sqrt{2}} \chi^{+} \\
\end{pmatrix} $ & $\begin{pmatrix} 1,3, 1 \end{pmatrix}$
\\ \\
  &  
& $\xi = \begin{pmatrix}
\xi^+ \\ \xi^0 \\ \xi^-
\end{pmatrix}$  & $\begin{pmatrix} 1,3, 0 \end{pmatrix}$
\\ \\ \hline  
 & SM and EW$\nu_R^M$ gauge fields &  $SU(3)_c \otimes SU(2)_W \otimes U(1)_Y$ &
 \\ \hline \\
& $W^a_\mu$  & $\begin{pmatrix} 1,3, 0 \end{pmatrix}$ 
&   
\\ \\
& $B_\mu$  & $\begin{pmatrix} 1,1, 0 \end{pmatrix}$ 
&   
\\ \\
& $g^b_\mu$  & $\begin{pmatrix} 8,1, 0 \end{pmatrix}$ 
&   
\\ \\ \hline 
\end{tabular}
}
\end{center}
\caption{\small Elementary fields in the commutative EW$\nu_R^M$ model. Each symbol $\nu_e$, $e$, $u$, or $d$ stands for the three corresponding SM flavors. There is no mixing of mirror leptons and no mixing of mirror quarks.}
\label{ewrhn-table}
\end{table}

The corresponding $\mathcal{L}_{EW\nu_R^M}$ Lagrangian constructed by Hung \cite{EWRH-Hung-2007} on the Minkowski spacetime is composed of the SM Lagrangian supplemented by, the kinetic term for the mirror fermions $\mathcal{L}_{Fermions}^{M}$, the kinetic terms and the potential for the new scalar fields expressed in  $\mathcal{L}_{Higgs}^{M}$, the interaction between the standard model scalar and the new scalar fields in $\mathcal{L}_{Higgs}^{M-SM}$, the Yukawa interactions for the mirror fermions $\mathcal{L}_{Yukawa}^{M}$, and the combined mirrors and SM fermions Yukawa interactions $\mathcal{L}_{Yukawa}^{M-SM}$. Hence we can write: 
\begin{equation}
\mathcal{L}_{EW\nu_R^M} = \mathcal{L}_{Gauge}^{SM} + \mathcal{L}_{Fermions} + \mathcal{L}_{Higgs} + \mathcal{L}_{Yukawa} +\mathcal{L}_{GF} + \mathcal{L}_{FP} \, ,
\end{equation}
where the kinetic Lagrangian for the fermions is:
\begin{equation}
\mathcal{L}_{Fermions} = \mathcal{L}_{Fermions}^{SM} +\mathcal{L}_{Fermions}^{M} \, ,
\end{equation}
with $\mathcal{L}_{Fermions}^{M}$ having for the mirror fermions, an expression similar to that of the Lagrangian $\mathcal{L}_{Fermions}^{SM}$ for the SM fermions with the difference that the SM left handed doublet fermions are replaced by the right handed doublet mirror fermions and the SM right handed singlet fermions are replaced by the left handed singlet mirror fermions. 

The scalar sector is governed by the Lagrangian \cite{EWRH-Hung-2016}:
\begin{eqnarray} \nonumber
\mathcal{L}_{Higgs} &=& \frac{1}{2} Tr\left[ \left(D_\mu \Phi_{2} \right)^\dagger \left(D^\mu \Phi_{2}\right)\right]  \\ \nonumber
&&
+ \frac{1}{2} \left\{ \vert\partial_\mu \varphi_{_S} \vert^2 + Tr\left[ \left(D_\mu \Phi_{2M} \right)^\dagger \left(D^\mu \Phi_{2M}\right) \right] + Tr\left[ \left(D_\mu X \right)^\dagger \left(D^\mu X \right) \right]\right\} \\
&& + V(\varphi_s, \Phi_2, \Phi_{2M}, X) = \mathcal{L}_{Higgs}^{SM} + \mathcal{L}_{Higgs}^{M} +\mathcal{L}_{Higgs}^{M-SM} \, ,
\end{eqnarray}
with $V(\varphi_s, \Phi_2, \Phi_{2M}, X)$ being the potential interaction between the scalar fields \cite{EWRH-Hung-2016},
and $X$ is the $3 \times 3$ matrix representation of the two scalar triplets written as follow \cite{GM-1985}:
\begin{equation}
X = \begin{pmatrix}
{\chi^{0}}^* & \xi^+ & \chi^{++} \\
\chi^- & \xi^0 & \chi^+ \\
\chi^{--}& \xi^- & \chi^{0}
\end{pmatrix}\,,
\end{equation}
with $\chi^{--} = {\chi^{++}}^*$, $\chi^{-} = -{\chi^{+}}^*$, $\xi^{-} = - {\xi^{+}}^*$, and the covariant derivatives acting on the scalar sector are given by \cite{EWRH-Hung-2016,GM-1985,Hung-2013}:
\begin{eqnarray}
\left\{\begin{array}{l}
D_\mu \Phi_2 \equiv \partial_\mu \Phi_2 + i\dfrac{g}{2} (\overrightarrow{W}_\mu.\vec{\tau}) \Phi_2 + i \dfrac{g^\prime}{2} B_\mu \Phi_2 \,,\\ 
D_\mu \Phi_{2M} \equiv \partial_\mu \Phi_{2M }+ i\dfrac{g}{2} (\overrightarrow{W}_\mu.\vec{\tau}) \Phi_{2M} + i \dfrac{g^\prime}{2} B_\mu \Phi_{2M} \,,\\ 
D_\mu X \equiv \partial_\mu X + i{g} (\overrightarrow{W}_\mu.\overrightarrow{T}) X - i {g^\prime} B_\mu X T_3 \,,\\ 
\end{array}
\right.
\end{eqnarray}
where $\tau_i$ are the Pauli matrices, and the $T_i$ are the following 
$3\times 3$ matrix representation of the $SU(2)$ generators:
\begin{equation}
    T_1 = \frac{1}{\sqrt{2}}\begin{pmatrix}
        0 & 1 & 0 \\
        1 & 0 & 1 \\
        0 & 1 & 0
    \end{pmatrix} \,,\qquad
    T_2 = \frac{1}{\sqrt{2}}\begin{pmatrix}
        0 & -i & 0 \\
        i & 0 & -i \\
        0 & i & 0
    \end{pmatrix} \,,\qquad
    T_3 = \begin{pmatrix}
        1 & 0 & 0 \\
        0 & 0 & 0 \\
        0 & 0 & -1
    \end{pmatrix}\,.
\end{equation}
The remaining $\mathcal{L}_{Yukawa} $ that enables fermions to acquire masses through electroweak symmetry breaking is~:
\begin{equation}\label{equ-Yukawa}
\mathcal{L}_{Yukawa} = \mathcal{L}_{Yukawa}^{SM} +\mathcal{L}_{Yukawa}^{M} +\mathcal{L}_{Yukawa}^{M-SM}   \,,
\end{equation}
where $\mathcal{L}_{Yukawa}^{SM}$ is the standard model Yukawa Lagrangian, and
\begin{equation} \label{equ-Yukawa-mirror}
\mathcal{L}_{Yukawa}^{M} = \mathcal{L}_{e^M} + \mathcal{L}_{q^M} + \mathcal{L}_{\nu_{e_R}},
\end{equation} 
is the pure mirror Yukawa's interaction with \cite{EWRH-Hung-2014}:
\begin{eqnarray} \label{equ-Yukawa-eM}
\mathcal{L}_{e^M} &=& -g_{{e^M}} \bar{l}_R^M \Phi_{2M} e_L^M - g_{{e^M}} \bar{e}_L^M \Phi_{2M}^\dagger l_R^M  \,,\\
\label{equ-Yukawa-qM}
\mathcal{L}_{q^M} &=& - g_{{d^M}} \bar{q}_R^M \Phi_{2M} d_L^M  - g_{{d^M}} \bar{d}_L^M \Phi_{2M}^\dagger q_R^M - g_{{u^M}} \bar{q}_R^M \tilde{\Phi}_{2M} u_L^M - g_{{u^M}} \bar{u}_L^M \tilde{\Phi}_{2M}^\dagger q_R^M \,, 
\\ \label{equ-Yukawa-nuR}
\mathcal{L}_{\nu_{e_R}}  &=& g_{_M} l_R^{M,T} \sigma_2 \tau_2 \tilde{\chi} l_R^M \,.
\end{eqnarray}
In the previous Equations (\ref{equ-Yukawa-eM}), (\ref{equ-Yukawa-qM}) and (\ref{equ-Yukawa-nuR}) we set $\tilde{\Phi}_{2M} = i \tau_2 \Phi_{2M}^{*}$, and we assume universal couplings $g_{_M}$, $g_{{e^{M}}}$, and $g_{{q^M}} = g_{{u^M}} = g_{{d^M}}$ whatever is the flavor. 
Moreover, a global symmetry is introduced to prevent Majorana mass term for left handed neutrinos \cite{EWRH-Hung-2007}.
Besides, we also have the Yukawa interaction that mixes between mirror and SM fermions :
\begin{equation}
\mathcal{L}_{Yukawa}^{M-SM} = \mathcal{L}_{S_l} +\mathcal{L}_{S_q},
\end{equation}
where :
\begin{eqnarray}
\mathcal{L}_{S_l}  &=& - g_{_{S_l}} \left( \bar{l}_{_L} l_{_R}^M  + \bar{l}_{_R}^M l_{_L} \right) \varphi_{_S} \,,  \\ 
\mathcal{L}_{S_q} &=& - g_{_{S_q}} \left( \bar{q}_{_R}^M q_{_L} +  \bar{q}_{_L} q_{_R}^M \right) \varphi_{_S}  - g^\prime_{_{S_q}} \left( \bar{q}_{_L}^M q_{_R} + \bar{q}_{_R} q_{_L}^M \right) \varphi_{_S} \, .
\end{eqnarray}
Finally, we draw the reader's attention to the fact that the gauge fixing Lagrangian~$\mathcal{L}_{GF}$ and the ghosts Lagrangian~$\mathcal{L}_{FP}$ keep the same expressions as those of the standard model~\cite{Aoki-82, Muta-2010} but with the three Goldstone fields expressed as a combination of the components of all the doublets and triplets scalar fields involved in this extension as in \cite{GM-1985, Hung-2013}.
\subsection{The mirror Yukawa Euclidean noncommutative-EW\texorpdfstring{$\nu_R^M$}{} Lagrangian}
Starting from the Minkowski commutative Yukawa Lagrangian $\mathcal{L}_{Yukawa}^{\scriptscriptstyle M}$, to construct the mirror Yukawa Euclidean noncommutative EW$\nu_R^M$ Lagrangian $\mathcal{L}_{Yukawa}^{\scriptscriptstyle M,ENC}$, we follow the following two steps\footnote{The construction of the whole EW$\nu_R^M$ theory on noncommutative spacetime will require in addition deforming the covariant derivatives \cite{TK-2020,Blaschke_2008} and is left for future work.} 
 :
\begin{enumerate}
    \item[i)] First, we transform the $\mathcal{L}_{Yukawa}^{M}$ Lagrangian from the Minkowski to the Euclidean commutative spacetime, by applying the Wick rotation to get $\mathcal{L}_{Yukawa}^{M,(Eucl.)}$ which simply includes a relative minus sign. 
    \item[ii)] The second step consists of two actions: 
\begin{itemize}
\item  First, we replace in the commutative $\mathcal{L}_{Yukawa}^{M,(Eucl.)}$, the ordinary product between fields by the star Weyl-Moyal product \cite{Moyal-product-2008, A-G_V-M_2003, HKM_2004} that enables us to properly incorporate the effect of the noncommutative spacetime, and is defined as follow ($\hbar= c = 1$): 
\begin{equation}
f_1(x) \star f_2(x) =f_1(x) \left[ e^{\frac{i}{2}\theta^{\mu \nu }\frac{\overleftarrow{\partial }}{\partial x_\mu } 
\frac{\overrightarrow{\partial }}{\partial y_\nu }}\right] f_2(y)\vert_{x=y} \,,
\end{equation}
where $f_1$ and $f_2$ stand for fields and $x$ and $y$ are the spacetime positions with :
\begin{equation}
x^{\mu }\star x^{\nu }-x^{\nu }\star x^{\mu }=\left[ x^{\mu }, x^{\nu }\right]_\star =i\theta ^{\mu \nu } \,,
\end{equation}
and in the Euclidean spacetime the $\theta ^{\mu \nu }$ are the components of the following totally antisymmetric tensor : 
\begin{equation}
\left[ \theta ^{\mu \nu }\right] =%
\begin{bmatrix}
0 & \theta & 0 & 0 \\ 
-\theta & 0 & 0 & 0 \\ 
0 & 0 & 0 & \theta \\ 
0 & 0 & -\theta & 0
\end{bmatrix} \, ,
\end{equation}
in which $\theta$ is the noncommutative small real deformation parameter and has a dimension inverse to the square of the energy.
\item Secondly, we add all permutations between fields and use the trace property of the star product~\cite{Bouchachia-2015}, which ultimately leads to the following substitution when going from the commutative Yukawa interaction between a spinor field $\psi(x)$ and a scalar field $\phi(x)$ to the corresponding noncommutative ones:
\begin{equation}
    g^{(c)} \bar{\psi}(x) \psi(x) \phi(x) \to g^{(nc)} \bar{\psi}(x) \star \psi(x) \star \phi(x) + \tilde{g}^{(nc)} \bar{\psi}(x) \star \phi(x) \star \psi(x) \,,
\end{equation}
where upscript $c$ stands for commutative and upscript $nc$ for noncommutative. Since in what follows, we are dealing with noncommutative cases, these indices will be dropped in the subsequent, except in a few possible confusing cases. 
\end{itemize}
\end{enumerate}
\vspace{-0.2cm}
We can summarize our procedure as follows:
\begin{equation}
\mathcal{L}_{Yukawa}^{\scriptscriptstyle M} \quad \xrightarrow[]{\text{Wick rotation}} \quad \mathcal{L}_{Yukawa}^{\scriptscriptstyle M, (Eucl.)} \quad \xrightarrow[]{\text{Moyal product}} \quad \mathcal{L}_{Yukawa}^{\scriptscriptstyle M,ENC} \,.
\end{equation}

Applying this procedure, and assuming universal noncommutative mirror Yukawa couplings respectively for all mirror down lepton's ($e^{M}$), all mirror quark's ($q^{M}$) and all mirror up lepton's ($\nu_{_{e_R}}$) enables us to write explicitly the Yukawa parts of the Lagrangian corresponding to each mirror fermionic sector as follows~:
\subsubsection*{\bf Mirror down leptons' Yukawa  sector : $\mathcal{L}_{e^{M}} \to \mathcal{L}_{e^{M}}^{\scriptscriptstyle ENC}$}
\begin{eqnarray}
\mathcal{L}_{e^{M}}^{\scriptscriptstyle ENC}\!\!\!\!\!&=&\!\!\!g_{e^{M}} \!\! \sum_{e.flavors} \!\! \left[ \overline{\nu}_{_{e_R}}^{\scriptscriptstyle M}
\star \phi_{2M}^{+} \star e_{L}^{M}+\overline{e}_{R}^{M} \star \phi_{2M}^{0} \star e_{L}^{M} - \overline{e}_{L}^M \star \phi_{2M}^{-} \star \nu_{_{e_R}}^{\scriptscriptstyle M} + \overline{e}_{L}^{M} \star \phi _{2M}^{0\,*} \star e_{R}^{M} \right] \nonumber \\
&& \!\!\!\!\!\!\! +\, \tilde{g}_{e^{M}} \!\!\! \sum_{e.flavors} \!\! \left[ \overline{\nu}_{_{e_R}}^{\scriptscriptstyle M} \star e_{L}^{M} \star \phi _{2M}^{+} + \overline{e}_{R}^{M} \star e_{L}^{M} \star \phi_{2M}^{0} - \overline{e}_{L}^M \star \nu_{_{e_R}}^{\scriptscriptstyle M} \star \phi_{2M}^{-} + \overline{e}_{L}^{M} \star e_{R}^{M} \star \phi_{2M}^{0\,*} \right] ,\nonumber \\
\end{eqnarray} 
with $\phi_{2M}^{-} = -\, \phi_{2M}^{+\,*}$.
\subsubsection*{\bf Mirror quarks' Yukawa sector : $\mathcal{L}_{q^{M}} \to \mathcal{L}_{q^{M}}^{\scriptscriptstyle ENC}$}
\begin{eqnarray}
\mathcal{L}_{q^{M}}^{\scriptscriptstyle ENC}\!\!\!\!\!&=&\!\!\!\!
g_{q^{M}} \!\! \sum_{q.flavors} \!\! \left[ \overline{u}_{R}^{M} \star \phi_{2M}^{+} \star d_{L}^{M} + \overline{d}_{R}^{M} \star \phi_{2M}^{0} \star d_{L}^{M} - \overline{d}_{L}^{M} \star \phi_{2M}^{-} \star u_{R}^{M} + \overline{d}_{L}^{M} \star \phi_{2M}^{0} \star d_{R}^{M} \right. \nonumber \\
&&
\left. \quad\quad + \, \overline{u}_{R}^{M} \star \phi_{2M}^{0\,*} \star u_{L}^{M} + \overline{d}_{R}^{M} \star \phi_{2M}^{-} \star
u_{L}^{M} + \overline{u}_{L}^{M} \star \phi_{2M}^{0} \star u_{R}^{M} - \overline{u}_{L}^{M} \star \phi_{2M}^{+} \star d_{R}^{M} \right] \nonumber \\
&&\!\!\!\!\!\! +\, \tilde{g}_{q^{M}} \!\!\!\!\sum_{q.flavors} \!\!\! \left[ \overline{u}_{R}^{M} \star d_{L}^{M} \star \phi_{2M}^{+} + \overline{d}_{R}^{M} \star d_{L}^{M} \star \phi_{2M}^{0} - \overline{d}_{L}^{M} \star u_{R}^{M} \star \phi_{2M}^{-} + \overline{d}_{L}^{M} \star d_{R}^{M} \star \phi _{2M}^{0\,*} \right. \nonumber \\
&& \left. \quad\quad + \, \overline{u}_{R}^{M} \star u_{L}^{M} \star \phi_{2M}^{0\,*} + \overline{d}_{R}^{M} \star u_{L}^{M} \star \phi_{2M}^{-} + \overline{u}_{L}^{M} \star u_{R}^{M} \star \phi_{2M}^{0} - \overline{u}_{L}^{M} \star d_{R}^{M} \star \phi_{2M}^{+} \right] . \nonumber \\
\end{eqnarray}
\subsubsection*{\bf Majorana right handed neutrinos' Yukawa sector : $\mathcal{L}_{\nu_{_{e_R}}} \to \mathcal{L}_{\nu_{_{e_R}}}^{\scriptscriptstyle ENC}$}
\begin{eqnarray}
\mathcal{L}_{\nu_{_{e_R}}}^{\scriptscriptstyle ENC} \!\!\!\! &=& \!\! g_{_M} \!\! \sum_{e.flavors} \!\! \left( \nu_{_{e_R}}^{\scriptscriptstyle M, T}\star \chi ^{0}\star i\sigma_{2}\,\nu_{_{e_R}}^{\scriptscriptstyle M}-\frac{1}{\sqrt{2}}\nu_{_{e_R}}^{\scriptscriptstyle M, T}\star \chi ^{+}\star i\sigma_{2}\,e_{_R}^{\scriptscriptstyle M}-\frac{1}{\sqrt{2}}e_{_R}^{ \scriptscriptstyle M,T}\star \chi ^{+}\star i\sigma_{2}\,\nu_{_{e_R}}^{\scriptscriptstyle M} \right. \nonumber \\
&& \bigg. \!\!\! -e_{_R}^{\scriptscriptstyle M,T}\star \chi ^{++}\star i\sigma _{2}\,e_{_R}^{\scriptscriptstyle M} \biggr)  + \! \widetilde{g}_{_M} \!\! \sum_{e.flavors} \!\! \left(\nu_{_{e_R}}^{\scriptscriptstyle M, T}\star i\sigma _{2}\,\nu_{_{e_R}}^{\scriptscriptstyle M}\star \chi ^{0}-\frac{1}{\sqrt{2}}\nu_{_{e_R}}^{\scriptscriptstyle M,T}\star i\sigma _{2}\,e_{_R}^{\scriptscriptstyle M}\star \chi ^{+} \right. \nonumber \\
&& \bigg. \!\!\! -\frac{1}{\sqrt{2}} e_{_R}^{\scriptscriptstyle M,T}\star i\sigma _{2}\,\nu_{_{e_R}}^{\scriptscriptstyle M}\star \chi ^{+} - e_{_R}^{\scriptscriptstyle M,T}\star i\sigma_{2}\,e_{_R}^{\scriptscriptstyle M}\star \chi ^{++} \biggr)\! .
\end{eqnarray}
It is worth noting here that in this transition from commutative to noncommutative formulations, the Euclidean two points scalar and spinor functions have to undergo the vulcanisation procedure \cite{Gurau-2009, Bouchachia-2015} to prevent subsequent UV/IR mixing and gain renormalizability.
\section{Renormalization of the mirror Yukawa sector}
\label{section-3}
\subsection{Renormalization constants}
The bare quantities\footnote{In previous section all couplings and fields were bare ones.} $\psi_{_0}$, $\phi_{_0}$, $g_{_{0,X}}$ and $\tilde{g}_{_{0,X}}$, and renormalized quantities $\psi_r$, $\phi_r$, $g_{_X,r}$ and $\tilde{g}_{_X,r}$ are related through the renormalization constants $Z_\psi$, $Z_\phi$, $Z_{g_{_X}}$ and $Z_{\tilde{g}_{_X}}$ as follows :

\begin{equation}
\psi_{_0} =\sqrt{Z_{\psi }} \, \psi_r, 
\quad\quad  \phi_{_0} =\sqrt{Z_{\phi }} \, \phi_r, 
\quad\quad  g_{_{0,X}}=Z_{g_{_X}} \, g_{_X,r},
\quad\quad \mathrm{and} \,\quad \tilde{g}_{_{0,X}}=Z_{\tilde{g}_{_X}} \, \tilde{g}_{_X,r},
\end{equation}
where, depending on the case, the index $X$ refers to $M,\, e^{M}$ or $ q^{M}$. The spinor field $\psi$ stands for
$\psi \equiv \nu_{e_{\scriptscriptstyle R}}^{\scriptscriptstyle M}, e_{\scriptscriptstyle L}^{\scriptscriptstyle M}, e_{\scriptscriptstyle R}^{\scriptscriptstyle M}, u_{\scriptscriptstyle L}^{\scriptscriptstyle M}, u_{\scriptscriptstyle R}^{\scriptscriptstyle M}, d_{\scriptscriptstyle L}^{\scriptscriptstyle M}, d_{\scriptscriptstyle R}^{\scriptscriptstyle M}$; 
with $e$, $u$, and $d$ representing each, respectively, the three mirror down lepton flavors, the three mirror up quark flavors, and the three mirror down quark flavors.
Whereas the scalar field $\phi$ stands for:
$\phi \equiv \chi^{\scriptscriptstyle 0}, \chi^{+}, \chi^{++}, \phi_{\scriptscriptstyle 2M}^{\scriptscriptstyle 0}, \phi_{\scriptscriptstyle 2M}^{+}, \phi_{\scriptscriptstyle 2M}^{-}$ \footnote{We do not consider here the $\xi$ and $\varphi_s$ fields since they are not involved in the purely mirror Yukawa sector with which we are dealing.}.

For simplicity, we assume that the scalar field's renormalization constants fulfill the following relations :
\begin{equation}
\begin{array}{l}
\sqrt{Z_{\chi }}\equiv \sqrt{Z_{\chi^{0}}}=\sqrt{Z_{\chi^{+}}}=\sqrt{Z_{\chi^{++}}} \,\,\, ,\\
\sqrt{Z_{\phi }}\equiv \sqrt{Z_{\phi_{2M}^{0}}}=\sqrt{Z_{\phi_{2M}^{+}}}=\sqrt{Z_{\phi_{2M}^{-}}} \,\,\, .
\end{array}
\end{equation}
Furthermore, for the spinor field's renormalization constants, we assume that ${Z_{\nu_{e_R}^{\scriptscriptstyle M}}}$, $Z_{e_{R}^{M}}$, $Z_{e_{L}^{M}}$, $Z_{u_{R}^{M}}$, $Z_{u_{L}^{M}}$, $Z_{d_{R}^{M}}$ and $Z_{d_{L}^{M}}$ are invariant under flavor changing.
\vspace{-0.3cm}
\subsection{Slavnov Taylor Identities}
For the mirror down leptonic Yukawa sector, replacing the bare quantities in terms of the corresponding renormalized ones leads to the following form of the Lagrangian $\mathcal{L}_{e^{M}}^{\scriptscriptstyle ENC}$ :
\begin{eqnarray} \label{Lagrange-Slavnov-Taylor-lepton}
\mathcal{L}_{e^{M}}^{\scriptscriptstyle ENC} \!\!\!\!\!&=&
\!\!\!g_{e^{M}} \!\! \left[ Z_{\nu_{_R} e_{_L} \phi } \!\!\! \sum_{e.flavors} \!\!\!\! \left( \overline{\nu }_{e_R}^{\scriptscriptstyle M} \star \phi_{2M}^{+} \star e_{L}^{M} - \overline{e}_{L}^M \star \phi_{2M}^{-} \star \nu_{e_R}^{\scriptscriptstyle M} \right) + Z_{e_{_R} e_{_L} \phi} \!\!\! \sum_{e.flavors} \!\!\!\! \left( \overline{e}_{R}^{M} \star \phi _{2M}^{0} \star e_{L}^{M} \right. \right. \nonumber \\
&& + \Bigg. \left. \overline{e}_{L}^{M} \star \phi_{2M}^{0\,*} \star e_{R}^{M} \right) \Biggr]  
+ \tilde{g}_{e^{M}} \!\! \left[ \widetilde{Z}_{\nu_{_R} e_{_L} \phi} \!\!\! \sum_{e.flavors} \!\!\!\! \left( \overline{\nu}_{e_R}^{\scriptscriptstyle M} \star e_{L}^{M} \star \phi_{2M}^{+} - \overline{e}_{L}^M \star \nu_{e_R}^{\scriptscriptstyle M} \star \phi_{2M}^{-} \right) \right. \nonumber \\
&& \left. + \, \tilde{Z}_{e_{_R} e_{_L} \phi} \!\!\! \sum_{e.flavors} \!\!\!\! \left( \overline{e}_{R}^{M} \star e_{L}^{M} \star \phi_{2M}^{0} + \overline{e}_{L}^{M} \star e_{R}^{M} \star \phi_{2M}^{0\,*} \right) \right]\!\!,
\end{eqnarray}
where here the fields and couplings are the renormalized ones even if in all what follows, we drop the~$r$ index to not clutter the writing. The vertex renormalization constants $Z_{\nu_{_R} e_{_L} \phi}$, $Z_{e_{_R} e_{_L} \phi}$, $\tilde{Z}_{\nu_{_R} e_{_L} \phi}$ and $\tilde{Z}_{e_{_R} e_{_L} \phi}$ are connected to the field's and coupling's renormalization constants through the following relations:
\begin{equation}
\begin{array}{cc}
Z_{\nu_{_R} e_{_L} \phi}=Z_{g_{e^{M}}}\sqrt{Z_{\nu_{e_{_R}}^{\scriptscriptstyle M} }}\sqrt{Z_{e_{_L}^{\scriptscriptstyle M}}}\sqrt{Z_{\phi }} \,\,,
&\qquad Z_{e_{_R} e_{_L} \phi}=Z_{e_{_L} e_{_R} \phi} =Z_{g_{e^{M}}}\sqrt{Z_{e_{_R}^{\scriptscriptstyle M}}}\sqrt{Z_{e_{_L}^{\scriptscriptstyle M}}}\sqrt{Z_{\phi }} \,\,,
\\
{}\\
\tilde{Z}_{\nu_{_R} e_{_L} \phi }=Z_{\widetilde{g}_{e^{M}}}\sqrt{Z_{\nu_{e_{_R}}^{\scriptscriptstyle M} }}\sqrt{Z_{e_{_L}^{\scriptscriptstyle M}}}\sqrt{Z_{\phi }} \,\,, 
& \qquad \tilde{Z}_{e_{_R} e_{_L} \phi}= \tilde{Z}_{e_{_L} e_{_R} \phi} =Z_{\widetilde{g}_{e^{M}}}\sqrt{Z_{e_{_R}^{\scriptscriptstyle M}}}\sqrt{Z_{e_{_L}^{\scriptscriptstyle M}}}\sqrt{Z_{\phi }} \,\,.
\\
{}\\
\end{array}
\end{equation}
The universality of the mirror down leptonic Yukawa coupling constants leads to the following Slavnov-Taylor identities:
\begin{equation} \label{Slavnov-Taylor-leptons}
\begin{array}{l}
Z_{g_{e^{M}}}=\dfrac{Z_{\nu_{_R} e_{_L} \phi}}{\sqrt{Z_{\nu_{e_{_R}}^{\scriptscriptstyle M} }}\sqrt{Z_{e_{_L}^{\scriptscriptstyle M}}}\sqrt{Z_{\phi }}}=\dfrac{Z_{e_{_R} e_{_L} \phi}}{\sqrt{Z_{e_{_R}^{\scriptscriptstyle M}}}\sqrt{Z_{e_{_L}^{\scriptscriptstyle M}}}\sqrt{Z_{\phi }}}\,\,, \\
{}\\
Z_{\tilde{g}_{e^{M}}}=\dfrac{\widetilde{Z}_{\nu_{_R} e_{_L} \phi}}{\sqrt{Z_{\nu_{e_{_R}}^{\scriptscriptstyle M} }}\sqrt{Z_{e_{_L}^{\scriptscriptstyle M}}}\sqrt{Z_{\phi }}} =\dfrac{\widetilde{Z}_{e_{_R} e_{_L} \phi}
}{\sqrt{Z_{e_{_R}^{\scriptscriptstyle M}}}\sqrt{Z_{e_{_L}^{\scriptscriptstyle M}}}\sqrt{Z_{\phi }}} \,\,.
\end{array}
\end{equation}
Besides, for the mirror quark sector, the Lagrangian $\mathcal{L}_{q^{M}}^{\scriptscriptstyle ENC}$ expressed in terms of the renormalized quantities takes the form:
\begin{eqnarray} \nonumber
\mathcal{L}_{q^{M}}^{\scriptscriptstyle ENC}\!\!\!\!\! &=&\!\!\! g_{q^{M}} \!\!\! \sum_{q.flavors} \!\!\! \left[ Z_{u_{_R}d_{_L}\phi}  \left( \overline{u}_{R}^{M} \star \phi_{2M}^{+} \star d_{L}^{M} - \overline{d}_{L}^{M} \star \phi_{2M}^{-} \star u_{R}^{M} \right) + Z_{u_{_L}d_{_R}\phi} \left( \overline{d}_{R}^{M} \star \phi_{2M}^{-} \star u_{L}^{M}  \right. \right. \\ \nonumber
&& \left. \Big. \qquad \qquad -\,\overline{u}_{L}^{M} \star \phi_{2M}^{+} \star d_{R}^{M} \Bigr) \right.
+ \, Z_{d_{_R} d_{_L} \phi}  \left( \overline{d}_{R}^{M} \star \phi_{2M}^{0} \star d_{L}^{M} + \overline{d}_{L}^{M} \star \phi_{2M}^{0\,*} \star d_{R}^{M} \right) \\ \nonumber
&& \qquad \qquad \left.  +\, Z_{u_{_R} u_{_L} \phi} \Bigl( \overline{u}_{R}^{M} \star \phi_{2M}^{0\,*} \star u_{L}^{M} + \overline{u}_{L}^{M} \star \phi_{2M}^{0} \star u_{R}^{M} \Bigr) \right] \\ \nonumber
&& + \, \tilde{g}_{q^{M}} \!\!\!\! \sum_{q.flavors} \!\!\!\! \left[ \tilde{Z}_{u_{_R} d_{_L} \phi} \! \left( \overline{u}_{R}^{M} \star d_{L}^{M} \star \phi_{2M}^{+} - \overline{d}_{L}^{M} \star u_{R}^{M} \star \phi_{2M}^{-} \right)\! + \!\tilde{Z}_{u_{_L} d_{_R} \phi} \!\left( \overline{d}_{R}^{M} \star u_{L}^{M} \star \phi_{2M}^{-} \right. \right. \\
&& \left. \Big. \qquad \qquad - \,\overline{u}_{L}^{M} \star d_{R}^{M} \star \phi_{2M}^{+} \Bigr) \right.
+\tilde{Z}_{d_{_R} d_{_L} \phi} \left( \overline{d}_{R}^{M} \star d_{L}^{M} \star \phi_{2M}^{0} + \overline{d}_{L}^{M} \star d_{R}^{M} \star \phi_{2M}^{0\,*} \right) \nonumber \\ 
&& \left.  \qquad \qquad + \, \tilde{Z}_{u_{_R} u_{_L} \phi}  \left( \overline{u}_{R}^{M} \star u_{L}^{M} \star \phi_{2M}^{0\,*} + \overline{u}_{L}^{M} \star u_{R}^{M} \star \phi_{2M}^{0} \right) \right] \,.
\label{Lagrange-Slavnov-Taylor-quark}
\end{eqnarray}
The corresponding vertex renormalization constants $Z_{u_{_R} d_{_L} \phi}$, $Z_{d_{_R} d_{_L} \phi}$, $Z_{u_{_R} u_{_L} \phi}$, $\tilde{Z}_{u_{_R} d_{_L} \phi}$, $\tilde{Z}_{d_{_R} d_{_L} \phi }$ and $\tilde{Z}_{{u_{_R}} u_{_L} \phi }$ are connected to the field's and coupling's renormalization constants as follows :

\begin{equation}
\begin{array}{cc}
Z_{u_{_R}d_{_L}\phi }=Z_{g_{q^{M}}}\sqrt{Z_{u_{_R}^{\scriptscriptstyle M}}}\sqrt{Z_{d_{_L}^{\scriptscriptstyle M}}}\sqrt{Z_{\phi }} \,\,, 
& \quad 
Z_{d_{_R}d_{_L}\phi }= Z_{d_{_L}d_{_R}\phi } =Z_{g_{q^{M}}}\sqrt{Z_{d_{_R}^{\scriptscriptstyle M}}}\sqrt{Z_{d_{_L}^{\scriptscriptstyle M}}}\sqrt{Z_{\phi }}
 \,\,,  \\
{} \\
Z_{u_{_L}d_{_R}\phi }=Z_{g_{q^{M}}}\sqrt{Z_{u_{_L}^{\scriptscriptstyle M}}}\sqrt{Z_{d_{_R}^{\scriptscriptstyle M}}}\sqrt{Z_{\phi }} \,\,, 
& \quad 
Z_{{u_{_R}}u_{_L}\phi}= Z_{{u_{_L}}u_{_R}\phi} =Z_{g_{q^{M}}}\sqrt{Z_{u_{_R}^{\scriptscriptstyle M}}}\sqrt{Z_{u_{_L}^{\scriptscriptstyle M}}}\sqrt{Z_{\phi }}
 \,\,, \\
{} \\
\tilde{Z}_{u_{_R}d_{_L}\phi}=Z_{\tilde{g}_{q^{M}}}\sqrt{Z_{u_{_R}^{\scriptscriptstyle M}}}\sqrt{Z_{d_{_L}^{\scriptscriptstyle M}}}\sqrt{Z_{\phi }} \,\,,
& \quad \tilde{Z}_{d_{_R}d_{_L}\phi}= \tilde{Z}_{d_{_L}d_{_R}\phi} =Z_{\tilde{g}_{q^{M}}}\sqrt{Z_{d_{_R}^{\scriptscriptstyle M}}}\sqrt{Z_{d_{_L}^{\scriptscriptstyle M}}}\sqrt{Z_{\phi }} \,\,, \\
{} \\
\tilde{Z}_{u_{_L}d_{_R}\phi}=Z_{\tilde{g}_{q^{M}}}\sqrt{Z_{u_{_L}^{\scriptscriptstyle M}}}\sqrt{Z_{d_{_R}^{\scriptscriptstyle M}}}\sqrt{Z_{\phi }}
 \,\, , 
& \quad \tilde{Z}_{{u_{_R}}u_{_L}\phi}= \tilde{Z}_{{u_{_L}}u_{_R}\phi} =Z_{\tilde{g}_{q^{M}}}\sqrt{Z_{u_{_R}^{\scriptscriptstyle M}}}\sqrt{Z_{u_{_L}^{\scriptscriptstyle M}}}\sqrt{Z_{\phi }} \,\,,
\end{array}
\end{equation}
with the corresponding Slavnov-Taylor identities given by:
\begin{equation} \label{Slavnov-Taylor-quarks}
\left.
\begin{array}{l}
Z_{g_{q^{M}}}=\dfrac{Z_{u_{_R} d_{_L} \phi }}{\sqrt{Z_{u_{_R}^{\scriptscriptstyle M}}}\sqrt{Z_{d_{_L}^{\scriptscriptstyle M}}}\sqrt{Z_{\phi}}}
=\dfrac{Z_{d_{_R} d_{_L} \phi }}{\sqrt{Z_{d_{_R}^{\scriptscriptstyle M}}}\sqrt{Z_{d_{_L}^{\scriptscriptstyle M}}}\sqrt{Z_{\phi }}}
=\dfrac{Z_{{u_{_R}}u_{_L}\phi }}{\sqrt{Z_{u_{_R}^{\scriptscriptstyle M}}}\sqrt{Z_{u_{_L}^{\scriptscriptstyle M}}}\sqrt{Z_{\phi }}} \,\,,\\
Z_{\tilde{g}_{q^{M}}}
=\dfrac{\widetilde{Z}_{u_{_R}d_{_L} \phi}}{\sqrt{Z_{u_{_R}^{\scriptscriptstyle M}}}\sqrt{Z_{d_{_L}^{\scriptscriptstyle M}}}\sqrt{Z_{\phi }}}
=\dfrac{\widetilde{Z}_{d_{_R}d_{_L}\phi }}{\sqrt{Z_{d_{_R}^{\scriptscriptstyle M}}}\sqrt{Z_{d_{_L}^{\scriptscriptstyle M}}}\sqrt{Z_{\phi }}}
=\dfrac{\widetilde{Z}_{u_{_R}u_{_L} \phi }}{\sqrt{Z_{u_{_R}^{\scriptscriptstyle M}}}\sqrt{Z_{u_{_L}^{\scriptscriptstyle M}}} \sqrt{Z_{\phi }}} \,\,.
\end{array}
\right.
\end{equation}
Furthermore, for the Majorana right handed neutrinos' sector, in terms of renormalized quantities the Lagrangian $\mathcal{L}_{\nu_{e_R}}^{\scriptscriptstyle ENC}$ takes the form :
\begin{eqnarray} \nonumber
\mathcal{L}_{\nu_{e_R}}^{\scriptscriptstyle ENC}\!\!\!\!\!&=&\!\!\! g_{_M} \left[ Z_{\nu_{_R}^{T} \nu_{_R} \chi} \sum_{e.flavors}(\nu_{e_R}^{\scriptscriptstyle M,T}\star \chi^{0}\star i\sigma _{2}\, \nu_{e_R}) -Z_{e_{_R}^{T} e_{_R} \chi} \sum_{e.flavors} (e_{R}^{M,T}\star \chi ^{++}\star i\sigma_{2} \, e_{R}^{M}) \right. \\ \nonumber
&& \left. 
\!\!\!\!\! -\frac{1}{\sqrt{2}} \, Z_{\nu_{_R}^{T} e_{_R} \chi} \! \sum_{e.flavors} \!\!\! (\nu_{e_R}^{\scriptscriptstyle M,T}\star \chi^{+}\star i\sigma_{2}\, e_{R}^{M}) - \frac{1}{\sqrt{2}} \, Z_{\nu_{_R} e_{_R}^{T} \chi} \! \sum_{e.flavors} \!\!\! (e_{R}^{M,T}\star \chi^{+}\star i\sigma_{2}\,\nu_{e_R}^{\scriptscriptstyle M})
\right] \\ \nonumber
&& \!\!\!\!\! +\widetilde{g}_{_M} \left[ \tilde{Z}_{\nu_{_R}^{T} \nu_{_R} \chi} \sum_{e.flavors} (\nu_{e_R}^{\scriptscriptstyle M,T}\star i\sigma _{2} \, \nu_{e_R}^{\scriptscriptstyle M}\star \chi ^{0}) 
-\tilde{Z}_{e_{_R}^{T} e_{_R} \chi} \sum_{e.flavors} (e_{R}^{M,T} \star i\sigma_{2} \,e_{R}^{M} \star \chi^{++})
 \right. \\ 
&& \left. 
\!\!\!\!\!-\frac{1}{\sqrt{2}} \, \tilde{Z}_{\nu_{_R}^{T} e_{_R} \chi} \! \sum_{e.flavors} \!\!\! (\nu_{e_R}^{\scriptscriptstyle M,T}\star i\sigma_{2}\, e_{R}^{M} \star \chi ^{+}) -\frac{1}{\sqrt{2}} \, \tilde{Z}_{\nu_{_R} e_{_R}^{T} \chi} \! \sum_{e.flavors} \!\!\! ( e_{R}^{M,T} \star i\sigma_{2} \, \nu_{e_R}^{\scriptscriptstyle M} \star \chi^{+})
\right]\!.
\label{Lagrange-Slavnov-Taylor-neutrino}
\end{eqnarray}
In terms of the field's and coupling's renormalization constants, the vertex renormalization constants $Z_{\nu_{_R}^{T} \nu_{_R} \chi}$, $Z_{e_{_R}^{T} e_{_R} \chi}$, $Z_{\nu_{_R}^{T} e_{_R} \chi}$, $Z_{\nu_{_R} e_{_R}^{T} \chi}$, $\tilde{Z}_{\nu_{_R}^{T} \nu_{_R} \chi}$, $\tilde{Z}_{e_{_R}^{T} e_{_R} \chi}$, $\tilde{Z}_{\nu_{_R}^{T} e_{_R} \chi}$, and $\tilde{Z}_{\nu_{_R} e_{_R}^{T} \chi}$ can be written as :
\begin{equation}
\begin{array}{cc}
Z_{\nu_{_R}^{T} \nu_{_R} \chi}=Z_{g_{M}}\sqrt{Z_{\nu_{e_{_R}}^{\scriptscriptstyle M,T} }}\sqrt{Z_{\nu_{e_{_R}}^{\scriptscriptstyle M}} }\sqrt{Z_{\chi }} \,\,, 
& \quad \tilde{Z}_{\nu_{_R}^{T} \nu_{_R} \chi}=Z_{\tilde{g}_{M}}\sqrt{Z_{\nu_{e_{_R}}^{\scriptscriptstyle M,T} }}\sqrt{Z_{\nu_{e_{_R}}^{\scriptscriptstyle M} }}\sqrt{Z_{\chi }} \,\,, \\
{} \\
Z_{e_{_R}^{T} e_{_R} \chi}=Z_{g_{M}}\sqrt{Z_{e_{_R}^{\scriptscriptstyle M,T} }}\sqrt{Z_{e_{_R}^{\scriptscriptstyle M} }}\sqrt{Z_{\chi }} \,\,,
& \quad \tilde{Z}_{e_{_R}^{T} e_{_R} \chi}=Z_{\tilde{g}_{M}}\sqrt{Z_{e_{_R}^{\scriptscriptstyle M,T} }}\sqrt{Z_{e_{_R}^{\scriptscriptstyle M} }}\sqrt{Z_{\chi }} \,\,, \\
{} \\
Z_{\nu_{_R}^{T} e_{_R} \chi}=Z_{g_{M}}\sqrt{Z_{\nu_{e_{_R}}^{\scriptscriptstyle M,T} }}\sqrt{Z_{e_{_R}^{\scriptscriptstyle M} }}\sqrt{Z_{\chi }} \,\,,
& \quad \tilde{Z}_{\nu_{_R}^{T} e_{_R} \chi}=Z_{\tilde{g}_{M}}\sqrt{Z_{\nu_{e_{_R}}^{\scriptscriptstyle M,T} }}\sqrt{Z_{e_{_R}^{\scriptscriptstyle M} }}\sqrt{Z_{\chi }} \,\,, \\
{} \\
Z_{\nu_{_R} e_{_R}^{T} \chi}=Z_{g_{M}}\sqrt{Z_{\nu_{e_{_R}}^{\scriptscriptstyle M} }}\sqrt{Z_{e_{_R}^{\scriptscriptstyle M,T}}}\sqrt{Z_{\chi }} \,\,, 
& \quad \tilde{Z}_{\nu_{_R} e_{_R}^{T} \chi}=Z_{\tilde{g}_{M}}\sqrt{Z_{\nu_{e_{_R}}^{\scriptscriptstyle M} }}\sqrt{Z_{e_{_R}^{\scriptscriptstyle M,T}}}\sqrt{Z_{\chi }} \,\,, 
\end{array}
\end{equation}
with the corresponding Slavnov-Taylor identities expressed as:
\begin{equation} \label{Slavnov-Taylor-neutrinos}
\begin{array}{l}
Z_{g_{M}}
=\frac{Z_{\nu_{_R}^{T} \nu_{_R} \chi}}{\sqrt{Z_{\nu_{e_{_R}}^{\scriptscriptstyle M,T} }}\sqrt{Z_{\nu_{e_{_R}}^{\scriptscriptstyle M} }}\sqrt{Z_{\chi }}}
=\frac{Z_{e_{_R}^{T} e_{_R} \chi}}{\sqrt{Z_{e_{_R}^{\scriptscriptstyle M,T} }}\sqrt{Z_{e_{_R}^{\scriptscriptstyle M} }}\sqrt{Z_{\chi }}}
=\frac{Z_{\nu_{_R}^{T} e_{_R} \chi}}{\sqrt{Z_{\nu_{e_{_R}}^{\scriptscriptstyle M,T} }}\sqrt{Z_{e_{_R}^{\scriptscriptstyle M} }}\sqrt{Z_{\chi }}}
=\frac{Z_{\nu_{_R} e_{_R}^{T} \chi}}{\sqrt{Z_{\nu_{e_{_R}}^{\scriptscriptstyle M} }}\sqrt{Z_{e_{_R}^{\scriptscriptstyle M,T}}}\sqrt{Z_{\chi }}}  \,\,,\\
{}\\
Z_{\tilde{g}_{M}}
=\frac{\tilde{Z}_{\nu_{_R}^{T} \nu_{_R} \chi}}{\sqrt{Z_{\nu_{e_{_R}}^{\scriptscriptstyle M,T} }}\sqrt{Z_{\nu_{e_{_R}^{\scriptscriptstyle M}} }}\sqrt{Z_{\chi }}}
=\frac{\tilde{Z}_{e_{_R}^{T} e_{_R} \chi}}{\sqrt{Z_{e_{_R}^{\scriptscriptstyle M,T} }}\sqrt{Z_{e_{_R}^{\scriptscriptstyle M} }}\sqrt{Z_{\chi }}}
=\frac{\tilde{Z}_{\nu_{_R}^{T} e_{_R} \chi}}{\sqrt{Z_{\nu_{e_{_R}}^{\scriptscriptstyle M,T} }}\sqrt{Z_{e_{_R}^{\scriptscriptstyle M} }}\sqrt{Z_{\chi }}}
=\frac{\tilde{Z}_{\nu_{_R} e_{_R}^{T} \chi}}{\sqrt{Z_{\nu_{e_{_R}}^{\scriptscriptstyle M} }}\sqrt{Z_{e_{_R}^{\scriptscriptstyle M,T}}}\sqrt{Z_{\chi }}}\,\,.
\end{array}
\end{equation}
Equipped with the previous Slavnov-Taylor identities (\ref{Slavnov-Taylor-leptons}), (\ref{Slavnov-Taylor-quarks}) and (\ref{Slavnov-Taylor-neutrinos}), to determine for each mirror Yukawa coupling the corresponding expression of the renormalization constant, it suffices to choose from the right-hand side of these identities one of the given expressions and to use it as a basis for calculation. 
Thus, under this choice, for each mirror Yukawa coupling, we keep only one term in the underlying renormalized mirror Yukawa Lagrangian, draw the Feynman diagrams of the corresponding process at the loop level, and then compute the necessary radiative corrections that lead to the desired $\beta$ functions.
\subsection{\texorpdfstring{$\beta$}{} function}
The dependence on the renormalization scale $\mu$ of a renormalized dimensionless coupling constant $g$ is given through the $\beta$ function defined by: 
\begin{equation}
    \beta_{g} = \mu \frac{dg}{d\mu} = \frac{dg}{dt} \,\,,
\end{equation}
where $t$ is the dimensionless quantity that we relate to the renormalization scale $\mu$ through the relation $t= \ln(\mu/m_t)$, with $m_t$ being the mass of the top quark.

In this work, we are interested in looking for the dependence of the mirror Yukawa couplings on the renormalization scale in an Euclidean noncommutative spacetime of dimension $D$ which we have to bring to four at the end of the calculation.
Hence, to derive the corresponding mirror Yukawa $\beta$ functions from the belonging renormalization constants, we  use the following defining relation :
\vspace{-0.2cm}
\begin{equation} \label{beta-renormalization-cste}
\beta_{g^{\left(k\right)}}=
\underset{\varepsilon_{_{UV}} \rightarrow 0}{\lim }\,\mathit{\mu }\dfrac{\partial }{\partial \mathit{\mu }}g^{\left(k\right) }= 
\underset{\varepsilon_{_{UV}} \rightarrow 0}{\lim } \left\{ -\frac{\varepsilon_{_{UV}}}{2} g^{\left( k\right) }- g^{\left( k\right)} \,\dfrac{\mu }{Z_{g}^{(k)}}\dfrac{\partial Z_{g}^{(k)}}{\partial \mu}  \right\}=
-g^{\left( k\right) } \dfrac{\partial  \ln Z_{g}^{(k)}}{\partial \ln \mathit{\mu }} \,\,,
\end{equation}
with :
\begin{equation}
\begin{array}{r}
 Z_{g}^{(k)} =
\dfrac{Z_{\psi \psi^\prime \phi}^{(k)}}
{Z_{\psi }^{1/2} 
Z_{\psi^\prime }^{1/2} 
Z_{\phi }^{1/2} }, 
\qquad g^{\left( k\right) }=g,\tilde{g},
\quad \text{and}\quad \varepsilon_{_{UV}} = 4-D \,\,,
\end{array}
\end{equation}
and :
\begin{equation}
Z_{\psi \psi^\prime \phi }^{(k)}=1+\dfrac{\delta_{g}^{(k)}}{g^{\left( k\right) }},
\qquad Z_{\psi}=1+\delta_{\psi},
\qquad Z_{\psi^\prime }=1+\delta_{\psi^\prime},
\qquad Z_{\phi }=1+\delta_{\phi},
\end{equation}
where $\delta_{g}^{(k)}$ is the vertex counter term, and $\delta_{\psi}$, $\delta_{\psi^\prime}$, and $\delta_{\phi}$ are respectively spinor and scalar self energies counter terms.

Hence, we can write :
\vspace{-0.2cm}
\begin{equation} \label{beta-counter-terms}
\beta_{g^{\left(k\right)}}= g^{\left( k\right)}
\dfrac{\partial }{\partial \ln \mathit{\mu}}\left[-\,\dfrac{\delta_{g}^{(k)}}{g^{\left( k\right) }} + \dfrac{1}{2} \, \left( \delta _{\psi} +\,\delta _{\psi^\prime} \right) + \dfrac{1}{2} \, \delta _{\phi} \right].
\end{equation}

So to get the $\beta$ function that governs each of the mirror Yukawa couplings, we have to first select in respect of the Slavnov Taylor identities, one term from the corresponding mirror Yukawa Lagrangian. Then for a process that embodies this chosen term, we determine through radiative corrections calculation, the counter terms $\delta_g^{(k)}$, $\delta_\psi$, $\delta_{\psi^\prime}$ and $\delta_\phi$, needed for Equation (\ref{beta-counter-terms}).

For better clarity, and before getting into the details of the loop calculations, it's worthwhile to recall here, that at the high energy scale on which we are interested, our computational assumptions are:
\begin{itemize}
    \item The mirror Yukawa couplings $g_{_M}$, $g_{e^M}$, and $g_{q^M}$ are flavor independent.
    \item As in \cite{EWRH-Hung-2007}, relative to the mirror Yukawa couplings $g_{_M}$, $g_{e^M}$, and $g_{q^M}$, we neglect :
    \begin{enumerate}
        \item the gauge couplings $g$, $g^\prime$, and $g_s$,
        \item the $g_{_{S_l}}$, $g_{_{S_q}}$, and $g^\prime_{_{S_q}}$ couplings between mirror and standard model fermions, 
        \item the $\lambda_i$ couplings occurring in the $V(\varphi_s, \Phi_2, \Phi_{2M}, X)$ scalar potential.
    \end{enumerate}
    \item There is no mixing of mirror leptons and no mixing of mirror quarks.
\end{itemize}

\section{One loop counter terms for the mirror Yukawa sector}
\label{section-4}
To obtain the leading contributions to the $\beta$ functions for the mirror Yukawa couplings, we consider respectively the one loop radiative corrections of the process {\footnotesize $ \nu_{e_R}(p) + \nu_{e_R}^{T}(p^\prime) \to \chi^{0}(q)$} for the $g_{_M}$, $\tilde{g}_{_M}$ couplings, then of the process {\footnotesize $ e_{R}^{M}(p) + \overline{e}_{L}^{M}(p^\prime) \to \phi_{2M}^{0}(q)$} for the $g_{{e^M}}$, $\tilde{g}_{e^M}$ couplings, and then of the process {\footnotesize $u_{R}^{M}(p) + \overline{u}_{L}^{M}(p^\prime) \to \phi_{2M}^{0}(q)$} for the $g_{q^M}$ and $\tilde{g}_{q^M}$ couplings. The tree level Feynman diagrams belonging to these processes are depicted in Figure \ref{fig:Feynman-diagrams-tree-level}. 
The corresponding one loop Feynman diagrams are shown in the Appendix and are consistent with the assumptions highlighted at the end of the previous section.
\begin{figure}[ht]
     \centering
      \vspace{-0.4cm}
     \begin{subfigure}[b]{0.3\textwidth}
         \centering
\scalebox{.50}{ 
\fcolorbox{white}{white}{
  \begin{picture}(302,200) (133,-113)
    \SetWidth{1.0}
    \SetColor{Black}
    \Line[arrow,arrowpos=0.5,arrowlength=5,arrowwidth=2,arrowinset=0.2,clock](184,-102)(256,-42)
    \Line[arrow,arrowpos=0.5,arrowlength=5,arrowwidth=2,arrowinset=0.2](256,-42)(184,18)
    \Line[dash,dashsize=10](256,-42)(334,-42)
    \Text(196,15)[lb]{\Large{\Black{$ \bar{e}_L^M$}}}
    \Text(302,-36)[lb]{\Large{\Black{$\phi_{2M}^{0}$}}}
    \Text(196,-112)[lb]{\Large{\Black{$e_R^M$}}}
  \end{picture}
}}
         \caption{ \scriptsize $e_{R}^{M}(p) + \overline{e}_{L}^{M}(p^\prime) \to \phi_{2M}^{0}(q)$.}
         \label{fig:Feynman-Diagrams-tree-e-e-phi}
     \end{subfigure}
      \begin{subfigure}[b]{0.3\textwidth}
         \centering
\scalebox{.50}{ 
\fcolorbox{white}{white}{
  \begin{picture}(302,200) (133,-113)
    \SetWidth{1.0}
    \SetColor{Black}
    \Line[arrow,arrowpos=0.5,arrowlength=5,arrowwidth=2,arrowinset=0.2,clock](184,-102)(256,-42)
    \Line[arrow,arrowpos=0.5,arrowlength=5,arrowwidth=2,arrowinset=0.2](256,-42)(184,18)
    \Line[dash,dashsize=10](256,-42)(334,-42)
    \Text(196,15)[lb]{\Large{\Black{$ \bar{u}_L^M$}}}
    \Text(302,-36)[lb]{\Large{\Black{$\phi_{2M}^{0}$}}}
    \Text(196,-112)[lb]{\Large{\Black{$u_R^M$}}}
  \end{picture}
  }}
         \caption{\scriptsize $u_{R}^{M}(p) + \overline{u}_{L}^{M}(p^\prime) \to \phi_{2M}^{0}(q)$.}
         \label{fig:Feynman-Diagrams-tree-q-q-phi}
     \end{subfigure}
      \begin{subfigure}[b]{0.3\textwidth}
         \centering
\scalebox{.50}{ 
\fcolorbox{white}{white}{
  \begin{picture}(302,200) (133,-113)
    \SetWidth{1.0}
    \SetColor{Black}
    \Line[arrow,arrowpos=0.5,arrowlength=5,arrowwidth=2,arrowinset=0.2,clock](184,-102)(256,-42)
    \Line[arrow,arrowpos=0.5,arrowlength=5,arrowwidth=2,arrowinset=0.2](256,-42)(184,18)
    \Line[dash,dashsize=10](256,-42)(334,-42)
    \Text(196,15)[lb]{\Large{\Black{$ \nu_{e_R}^{M\,T}$}}}
    \Text(302,-36)[lb]{\Large{\Black{$\chi^{0}$}}}
    \Text(196,-112)[lb]{\Large{\Black{$\nu_{e_R}^M$}}}
  \end{picture}}}
         \caption{ \scriptsize $\nu_{e_R}^M(p) + \nu_{e_R}^{M\,T}(p^\prime) \to \chi^{0}(q)$.}
         \label{fig:Feynman-Diagrams-tree-nu-nu-phi}
     \end{subfigure}
              \caption[Caption for LOF]{\small Tree level Feynman diagrams for the processes used in the computation of the mirror Yukawa $\beta$ functions\protect\footnotemark .}
        \label{fig:Feynman-diagrams-tree-level}
\end{figure}
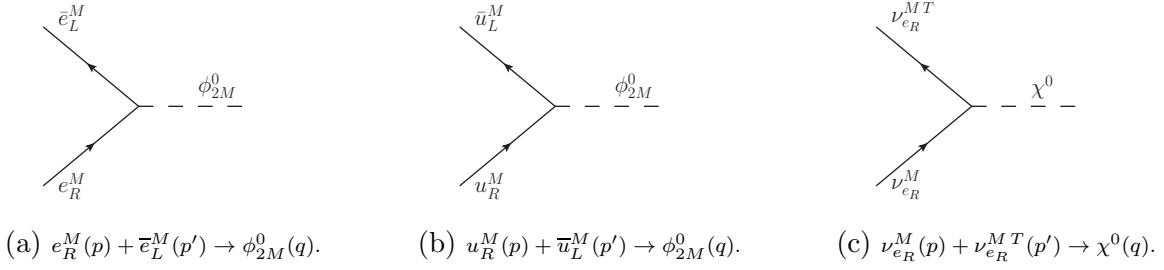
\footnotetext{Feynman diagrams are drawn using Jaxodraw\cite{Jaxodraw-2008}}

To achieve the calculation of the radiative corrections, we harness the work of Gurau et al~\cite{Gurau-2009} and Bouchachia et al  \cite{Bouchachia-2015} and adjust their results to our massless scalar and fermionic fields case's, to comply with the fact that we are dealing with a Lagrangian form formulated at energies before the occurrence of the electroweak symmetry breaking. Hence, in the Euclidean momentum space, for the scalar massless vulcanised propagator $\Delta (p)$, we use :
\vspace{-0.3cm}
\begin{equation} \label{vulcanised-scalar-propagator}
\begin{gathered}
\scalebox{.6}{
    \begin{picture}(180,56) (159,-139)
    \SetWidth{1.0}
    \SetColor{Black}
    \Line[dash,arrowpos=0.5,arrowlength=5,arrowwidth=2,arrowinset=0.2](160,-136)(320,-136)
    \Text(160,-130)[lb]{\Large{\Black{${\phi}$}}}
    \Text(304,-130)[lb]{\Large{\Black{$\phi$}}}
    \Line[arrow,arrowpos=1.0,arrowlength=5,arrowwidth=2,arrowinset=0.2](224,-120)(256,-120)
    \Text(240,-114)[lb]{\Large{\Black{$p$}}}
  \end{picture}
  }
\end{gathered} \quad = \qquad
\Delta (p)= \frac{1}{p^{2}+\dfrac{a^{2}}{\theta^2 p^{2}}} \,\,.
\end{equation}
And, for the spinor massless vulcanised propagator $S(p)$, we use :
\vspace{-0.3cm}
\begin{equation} \label{vulcanised-spinor-propagator}
\begin{gathered}
\scalebox{.6}{
    \begin{picture}(180,56) (159,-139)
    \SetWidth{1.0}
    \SetColor{Black}
    \Line[arrow,arrowpos=0.5,arrowlength=5,arrowwidth=2,arrowinset=0.2](160,-136)(320,-136)
    \Text(160,-130)[lb]{\Large{\Black{$\overline{\psi}$}}}
    \Text(304,-130)[lb]{\Large{\Black{$\psi$}}}
    \Line[arrow,arrowpos=1.0,arrowlength=5,arrowwidth=2,arrowinset=0.2](224,-120)(256,-120)
    \Text(240,-114)[lb]{\Large{\Black{$p$}}}
  \end{picture}
  }
\end{gathered} = \quad
S(p)=\dfrac{i\pslash+ib\dfrac{\tpslash}{\theta^2 p^{2}}}{p^{2}+\dfrac{b^{2}}{\theta^2 p^{2}}} \,\,,
\end{equation}
where $\theta$ is the dimensionful deformation parameter, $a$ and $b$ are real dimensionless vulcanised parameters and $\tpslash = \gamma_\mu \tilde{p}^\mu = \gamma_\mu \theta^{\mu\nu} p_\nu$.\\
For the Yukawa vertex $V_{g}\left(p^\prime, p \right)$, we use :
\vspace{-0.3cm}
\begin{equation}
\begin{gathered}
\scalebox{.53}{
 \fcolorbox{white}{white}{
  \begin{picture}(176,144) (147,-187)
    \SetWidth{1.0}
    \SetColor{Black}
    \Line[arrow,arrowpos=0.5,arrowlength=5,arrowwidth=2,arrowinset=0.2](176,-176)(240,-128)
    \Line[arrow,arrowpos=0.5,arrowlength=5,arrowwidth=2,arrowinset=0.2](240,-128)(176,-80)
    \Line[dash,dashsize=10,arrow,arrowpos=0.5,arrowlength=5,arrowwidth=2,arrowinset=0.2](240,-128)(304,-128)
    \Text(174,-125)[lb]{\Large{\Black{$p^\prime$}}}
    \Text(278,-160)[lb]{\Large{\Black{$q$}}}
    \Text(174,-150)[lb]{\Large{\Black{$p$}}}
    \Line[arrow,arrowpos=1.0,arrowlength=5,arrowwidth=2,arrowinset=0.2](176,-156)(192,-144)
    \Line[arrow,arrowpos=1.0,arrowlength=5,arrowwidth=2,arrowinset=0.2](192,-110)(176,-96)
    \Line[arrow,arrowpos=1.0,arrowlength=5,arrowwidth=2,arrowinset=0.2](272,-144)(288,-144)
    \Text(192,-182)[lb]{\Large{\Black{$\psi$}}}
    \Text(195,-90)[lb]{\Large{\Black{$\overline{\psi^{\,\prime}}$}}}
    \Text(288,-122)[lb]{\Large{\Black{$\phi$}}}
  \end{picture}
}
}
\end{gathered}
=
\quad
V_{g}\left(p^\prime, p \right) =-\left[ g\,e^{\frac{i}{2}p^\prime \tilde{p}}+\tilde{g}\,e^{-\frac{i}{2}p^\prime \tilde{p}}\right] \,\,,
\label{yukawa-vertex}
\end{equation}
where $p$ and $p^\prime$ stand for momenta, $p^\prime \tilde{p} = p^{\prime}_\mu\theta^{\mu\nu} p_\nu \equiv p^\prime \wedge p$, $q=p-p^\prime$, and $g$ and $\tilde{g}$ are the noncommutative Yukawa couplings related to the same vertex. $\psi$ and $\psi^\prime$ are spinor fields, and $\phi$ is a scalar field. 

To identify our needed counter terms, we use the minimal subtraction defining relations~\cite{Bouchachia-2015} that connect the divergent part of the radiative corrections to their corresponding counter terms. 

Hence :
\begin{itemize}
    \item  The relation between $\delta_{\phi}$ and $\delta_{a^{2}}$ counter terms, and the $\Pi_\phi^{(div)}$ divergent part of the self energy of the vulcanised massless scalar propagator observes the formula\footnote{$a_0$ and $b_0$ are bare parameters, while $a$ and $b$ are the corresponding renormalized ones.} :    
\begin{equation}  \label{conter-term-scalar-propagator}
\Pi_\phi^{(div)}  {-\delta _{\phi }p^{2}-\delta _{a^{2}}\frac{1}{\theta^2\,p^{2}}%
=\Pi_\phi^{(div)} -(Z_{\phi }-1)p^{2}-(a_{0}^{2}Z_{\phi }-a^{2})\frac{1}{\theta^2\,p^{2}}} =0.
\end{equation}
\item The relation between $\delta_{\psi }$ and $\delta_{b}$ counter terms, and the $\Sigma_\psi^{(div)}$ divergent part of the self energy of the vulcanised massless fermionic propagator fulfills the constraint:
\begin{equation} \label{conter-term-fermionic-propagator}
\Sigma_\psi^{(div)} + i\delta _{\psi }\psla-i\delta _{b}\frac{\tpsla}{\theta^2 \, p^{2}}
= \Sigma_\psi^{(div)} + i(Z_{\psi }-1)\psla -i(b_{0}Z_{\psi }-b)\frac{{\tpsla}}{\theta^2 \,p^{2}} =0.
\end{equation}
\item The relation between $\delta_{g}$ and $\delta_{\tilde{g}}$ counter terms, and the $\Gamma_{g}^{(div)}$ divergent part of the noncommutative Yukawa vertex is governed by:
\begin{equation} \label{conter-term-vertex}
\Gamma_{g}^{(div)} -\delta _{g}\,e^{\frac{i}{2}p\prime \tilde{p}}-\delta_{\tilde{g}}\,e^{-\frac{i}{2}p\prime \tilde{p}} 
=\Gamma_{g}^{(div)} + (Z_{\psi \psi \prime \phi }-1)e^{\frac{i}{2}p\prime \tilde{p}}+(\tilde{Z}_{\psi \psi \prime \phi }-1)e^{-\frac{i}{2}p\prime \tilde{p}} = 0.
\end{equation}
\end{itemize} 
So, to get the needed counter terms, we first use the one loop Feynman diagrams of Figures~\ref{fig:Feynman-diagrams-leptonic-corrections},~\ref{fig:Feynman-diagrams-quark-corrections},~and~\ref{fig:Feynman-diagrams-neutrino-corrections} drawn in the Appendix for each of the tree level processes depicted in Figure \ref{fig:Feynman-diagrams-tree-level}. Then, we apply the previous vulcanised Feynman rules and perform the integrals to get the corresponding radiative corrections (See Appendix for calculation details). Finally, we use Equations (\ref{conter-term-scalar-propagator}), (\ref{conter-term-fermionic-propagator}) and (\ref{conter-term-vertex}) to extract the following counter terms UV divergent parts :
\begin{enumerate}
    \item[i)] From  the process {\footnotesize $ e_{R}^{M}(p) + \overline{e}_{L}^{M}(p^\prime) \to \phi_{2M}^{0}(q)$}, we obtain for the mirror down lepton's Yukawa sector :
    \begin{itemize}
        \item  the vertex $\delta_{g_{e^{M}}}$ and $\delta_{\tilde{g}_{e^{M}}}$ counter terms :
\vspace{-0.2cm}
\begin{equation}  \label{equ-counter-term-gem-gemtilde}
\frac{\delta_{g_{e^{M}}}}{g_{e^{M}}}=\frac{\delta_{\tilde{g}_{e^{M}}}}{\tilde{g}_{e^{M}}}=-\dfrac{2\,g_{e^{M}} \, \tilde{g}_{e^{M}}}{\left( 4\pi \right)^{2}} \dfrac{1}{\varepsilon_{_{UV}} },
\end{equation}
\item the fermionic self energy $\delta_{e}$ counter term  :
\vspace{-0.2cm}
\begin{equation} \label{equ-counter-term-e}
\delta_{e}= \delta_{e_R^M} + \delta_{\tilde{e}_L^M} = \frac{3\left( g_{e^{M}}^{2}+\tilde{g}_{e^{M}}^{2}\right) +
\frac{3}{2}\left( g_{M}^{2}+\tilde{g}_{M}^{2}\right) }{\left( 4\pi\right) ^{2}}\frac{1}{\varepsilon_{_{UV}} },
\end{equation}
\item and the scalar self energy $\delta_{\phi^{0}}^{(e)} $ counter term  :
\vspace{-0.2cm}
\begin{equation} \label{equ-counter-term-phi0e}
\delta_{\phi^{0}}^{(e)}= 3 \times \frac{8\left( g_{e^{M}}^{2}+\tilde{g}_{e^{M}}^{2}\right) + 16\left( g_{q^{M}}^{2}+\tilde{g}_{q^{M}}^{2}\right) }{\left( 4\pi \right) ^{2}} \frac{1}{\varepsilon_{_{UV}} }.
\end{equation}
 \end{itemize}
\item[ii)] From the process: {\footnotesize $u_{R}^{M}(p) + \overline{u}_{L}^{M}(p^\prime) \to \phi_{2M}^{0}(q)$ } and by ignoring color charges, we infer for the mirror quark's Yukawa sector:
\begin{itemize}
    \item the vertex $\delta_{g_{q^{M}}}$ and $\delta_{\tilde{g}_{q^{M}}}$ counter terms :
\vspace{-0.2cm}
\begin{equation} \label{equ-counter-term-gqm-gqmtilde}
\frac{\delta_{g_{q^{M}}}}{g_{q^{M}}}=\frac{\delta_{\tilde{g}_{q^{M}}}}{\tilde{g}_{q^{M}}}=-\dfrac{2\, g_{q^{M}}\,\tilde{g}_{q^{M}}}{\left( 4\pi \right)^{2}}  \dfrac{1}{\varepsilon_{_{UV}} },
\end{equation}
\item the fermionic self energy $\delta_{q}$ counter term :
\vspace{-0.2cm}
\begin{equation} \label{equ-counter-term-q}
\delta_{q}= \delta_{u_R^M} +\delta_{\tilde{u}_L^M} = \frac{4\left[ g_{q^{M}}^{2}+\tilde{g}_{q^{M}}^{2}\right] }{
\left( 4\pi \right) ^{2}} \frac{1}{\varepsilon_{_{UV}} },
\end{equation}
\item and the scalar self energy $\delta_{\phi^{0}}^{(q)} $ counter term :
\vspace{-0.2cm}
\begin{equation} \label{equ-counter-term-phi0q}
\delta_{\phi^{0}}^{(q)}= \delta_{\phi^{0}}^{(e)} = 3\times \frac{8\left( g_{e^{M}}^{2}+\tilde{g}_{e^{M}}^{2}\right) + 16\left( g_{q^{M}}^{2}+\tilde{g}_{q^{M}}^{2}\right) }{\left( 4\pi \right) ^{2}} \frac{1}{\varepsilon_{_{UV}} }.
\end{equation}
\end{itemize}
\item[iii)] From the process {\footnotesize $ \nu_{e_R}(p) + \nu_{e_R}^{T}(p^\prime) \to \chi^{0}(q)$}, we get for the Majorana right handed neutrinos' Yukawa sector:
\begin{itemize}
    \item the vertex $\delta_{g_{_M}}$ and $\delta_{\tilde{g}_{_M}}$ counter terms :
\vspace{-0.2cm}
\begin{equation} \label{equ-counter-term-gm-gmtilde}
\frac{\delta_{g_{_M}}}{g_{_M}}=\frac{\delta_{\tilde{g}_{_M}}}{\tilde{g}_{_M}}=-\dfrac{2\,g_{_M} \, \tilde{g}_{_M}}{\left( 4\pi \right)^{2}}  \dfrac{1}{\varepsilon_{_{UV}} },
\end{equation}
\item the fermionic self energy $\delta_{\nu}$ counter term :
\vspace{-0.2cm}
\begin{equation} \label{equ-counter-term-nuR}
\delta_{\nu}=\delta_{\nu_{_R}} + \delta_{\nu_{_R}^{\scriptscriptstyle T}} = \frac{3\left( g_{_M}^{2}+\tilde{g}_{_M}^{2}\right)
+g_{e^{M}}^{2}+\tilde{g}_{e^{M}}^{2}}{ \left( 4\pi \right) ^{2}} \frac{1}{\varepsilon_{_{UV}} },
\end{equation}
\item and the scalar self energy $\delta_{\chi^{(0)}}^{(\nu)}$ counter term  :
\vspace{-0.2cm}
\begin{equation} \label{equ-counter-term-chi0-nuR}
\delta_{\chi^{(0)}}^{(\nu)}= 3 \times \frac{4 \left( g_{_M}^{2} + \tilde{g}_{_M}^{2}\right)}{ \left( 4\pi \right)^{2}} \frac{1}{\varepsilon_{_{UV}}}.
\end{equation}
\vspace{-0.5cm}
\end{itemize}
\end{enumerate}
It's worth drawing the reader's attention here to the fact that, for all our previously found UV divergent part counter terms, there is no explicit dependence on the noncommutative spacetime deformation parameter $\theta$ in agreement with the results found in \cite{GBRN-2017} for the noncommutative scalar QED. Also, there is no dependence on the vulcanisation parameters $a$ and $b$. This feature can be traced to the following decomposition  of the vulcanised scalar and spinor propagators\footnote{We use here as in \cite{BenGeloun_2008} the relationship: \[ \frac{1}{A+B} = \frac{1}{A} -\frac{1}{A}B\frac{1}{A+B}.\]} :

\begin{equation}
\Delta (p)= \frac{1}{p^{2}+\dfrac{a^{2}}{\theta^2 p^{2}}} = \frac{1}{p^2} - \frac{1}{p^2}\frac{a^2}{(\theta^2\,p^4 +a^2)},
\end{equation}
and
\begin{equation}
S(p)=\left( i \pslash + i b\frac{\tpslash}{\theta^2 p^2}\right) \frac{1}{p^{2}+\dfrac{b^{2}}{\theta^2 p^{2}}} = i\frac{\pslash}{p^2} + \frac{i b \tpslash}{\theta^2 p^4} - \frac{i b^2 \pslash}{p^2(\theta^2p^4 + b^2)} - \frac{ib^3 \tpslash}{\theta^2 p^4 (\theta^2 p^4 + b^2)}.
\end{equation}
When these propagators occur in the one loop integrals, only the first term of each of the corresponding decompositions contributes to the logarithmic UV divergent part. This term is free from the deformation parameter $\theta$ and the vulcanised parameters $a$ and $b$. All the remaining terms that depend on $\theta$, $a$, and $b$, have finite contributions to the loop integrals. Therefore, they do not contribute to the UV divergent parts of the counter terms in the minimal subtraction scheme of interest.

Lastly, it's worthwhile to point out the following :
\begin{enumerate}
    \item[i)]  First, when trying to get the commutative limit starting from the noncommutative expressions by letting $\theta \to 0$ in a direct, “naive” way, the vulcanised propagators no longer make sense. 
    In fact, to correctly obtain the commutative limit one should proceed as indicated by Magnen et al in \cite{Rivassau-2009}. Nevertheless, in this study, we are not concerned with the commutative limit, we instead focus on getting the most suitable link between the commutative and the noncommutative $\beta$ functions at the one loop order. 
    \item[ii)] Subsequently, by examining the formula (\ref{beta-counter-terms}) from which the $\beta$ functions are established, we notice that it depends on a sum involving the counter term $\delta_g^{(k)}$ of the Yukawa vertex and the counter terms~$\delta_\psi$, $\delta_{\psi^\prime}$, and $\delta_\phi$ of spinor and scalar propagators. As pointed out in \cite{Bouchachia-2015}, at the one loop order the link between commutative and noncommutative versions of the counter terms $\delta_g^{(k)}$ of Yukawa vertices expressed in terms of couplings follows a different formula than the link between commutative and noncommutative versions of the counter terms~$\delta_\psi$, $\delta_{\psi^\prime}$, and $\delta_\phi$ of the propagators expressed in terms of couplings. In fact, at the one loop order, starting from the commutative version $\beta_{g_{_X}^{(c)}}$ function expressed in the form of the formula (\ref{beta-counter-terms}) i.e. in term of the counter terms~$\delta_g^{(k)}$, $\delta_\psi$, $\delta_{\psi^\prime}$, and $\delta_\phi$ and the couplings $g^{(k)}$, to recover the corresponding noncommutative version $\beta_{g_{_X}}^{(nc)}$ $\left(\text{or respectively :}\,\, \beta_{\tilde{g}_{_X}}^{(nc)}\right)$, we have to perform the following substitutions :
    \begin{enumerate}
        \item[$\bullet$] For $g^{(k)}$, replace $g_{_X}^{(c)}$ by $g_{_X}$ $\left( \text{respectively by :}\,\, \tilde{g}_{_X} \right)$.
        \item[$\bullet$] In $\dfrac{\delta_g^{(k)}}{g^{(k)}}$ expression, replace $\left[g_{_X}^{(c)}\right]^2$  by $g_{_X} \tilde{g}_{_X}$ $\left( \text{respectively by :}\,\, \tilde{g}_{_X} g_{_X}  \right)$. 
        \item[$\bullet$] In $\delta_\psi + \delta_{\psi^\prime}$ and in $\delta_\phi$ expressions, replace $\left[g_{_X}^{(c)}\right]^2$ by $g_{_X}^2 + \tilde{g}_{_X}^2$
       $\left(\text{respectively by :} \,\, \tilde{g}_{_X}^2 + g_{_X}^2  \right)$.
    \end{enumerate}
    Therefore, given this substitution relation, there can be no exact proportionality between commutative and noncommutative $\beta$ functions for Yukawa mirror couplings. Moreover, since spinor and scalar propagators fulfill the same substitution formula, the more loops are involved in the radiative corrections of the propagators, the more the predominance of the propagator counter terms over the vertex counter terms becomes in favor of the propagator counter terms. The latter case, if fulfilled, induces an approximate proportionality rule between the commutative and noncommutative cases which would be governed by the counter terms of the propagators.
\end{enumerate}
\section{One loop noncommutative mirror Yukawa \texorpdfstring{$\beta$}{} functions}
\label{section-5}
Since in the $\beta$ function evaluation, we are concerned by renormalized dimensionless couplings $g_{_{X,Ren.}}$, while our counter terms are expressed in terms of couplings $g_{_{X,ren.}}$ carrying dimension, we first incorporate the connecting relation $g_{_{X,ren.}} = g_{_{X,Ren.}} \, \mu^{\varepsilon_{_{\scriptscriptstyle UV}}/2}$ to obtain the $\mu$ dependence of the counter terms. Then, we introduce the corresponding counter terms in the (\ref{beta-counter-terms}) formula of the $\beta$ function to get the following 
\begin{eqnarray} \label{Beta-function-equations-non-com}
\left\{\begin{array}{l}
\beta_{g_{_M}}^{(nc),dif}= \dfrac{d g_{_M}}{dt} =\dfrac{1}{32\,\pi^{2}} \left[ 15\, g_{_M}^{3}(t) + 15\, g_{_M}(t) \tilde{g}_{_M}^{2}(t)  + 4 \, g_{_M}^{2}(t) \tilde{g}_{_M}(t) +   g_{_M}(t) g_{e^{M}}^{2}(t) + g_{_M}(t) \tilde{g}_{e^{M}}^{2}(t)  \right],  \\ 
{} \\ 
\beta_{\tilde{g}_{_M}}^{(nc),dif}= \dfrac{d \tilde{g}_{_M}}{dt} =\dfrac{1}{32\,\pi^{2}}\left[ 15\,  \tilde{g}_{_M}^{3}(t) + 15\, g_{_M}^{2}(t) \tilde{g}_{_M}(t)  + 4 \, g_{_M}(t) \tilde{g}_{_M}^{2}(t) + \tilde{g}_{_M}(t) g_{e^{M}}^{2}(t)+ \tilde{g}_{_M}(t) \tilde{g}_{e^{M}}^{2}(t) \right], \\
{} \\
\beta_{g_{e^{M}}}^{(nc),dif}= \dfrac{d g_{e^{M}}}{dt} = \dfrac{1}{32\,\pi^{2}} \left[ 27\, g_{e^{M}}^{3}(t) + 27\, g_{e^{M}}(t) \tilde{g}_{e^{M}}^{2}(t) + 4 \, g_{e^{M}}^{2}(t) \tilde{g}_{e^{M}}(t)   \right. \\ {} \\
 \left. \qquad \qquad \qquad \qquad \qquad + \, \dfrac{3}{2} \, g_{e^{M}}(t) g_{{_M}}^{2}(t) +  \dfrac{3}{2} \, g_{e^{M}}(t) \tilde{g}_{{_M}}^{2}(t) + \, 48\, g_{q^{M}}^{2}(t) g_{e^{M}}(t) + 48 \, \tilde{g}_{q^{M}}^{2}(t) g_{e^{M}}(t) \right], \\
{}\\
\beta_{\tilde{g}_{e^{M}}}^{(nc),dif}= \dfrac{d \tilde{g}_{e^{M}}}{dt} = \dfrac{1}{32\,\pi^{2}} \left[27 \, \tilde{g}_{e^{M}}^{3}(t) + 27 \, g_{e^{M}}^{2}(t) \tilde{g}_{e^{M}}(t) + 4 \, g_{e^{M}}(t) \tilde{g}_{e^{M}}^{2}(t)  
\right. \\ {} \\
 \left. \qquad \qquad \qquad \qquad \qquad + \, \dfrac{3}{2} \, \tilde{g}_{e^{M}}(t) g_{{_M}}^{2}(t) + \dfrac{3}{2} \tilde{g}_{e^{M}}(t) \tilde{g}_{{_M}}^{2}(t) + \, 48\, g_{q^{M}}^{2}(t) \tilde{g}_{e^{M}}(t) + 48 \, \tilde{g}_{q^{M}}^{2}(t) \tilde{g}_{e^{M}}(t) \right], \\
{} \\
\beta_{g_{q^{M}}}^{(nc),dif}= \dfrac{d g_{q^M}}{dt} =\dfrac{1}{8\,\pi^{2}} \Bigl[ 13\, g_{q^{M}}^{3}(t) + 13\, g_{q^{M}}(t) \tilde{g}_{q^{M}}^{2}(t)  + g_{q^{M}}^{2}(t) \tilde{g}_{q^{M}}(t)  
+ 6\, g_{q^{M}}(t) g_{e^{M}}^{2}(t) + \,6\, g_{q^{M}}(t) \tilde{g}_{e^{M}}^{2}(t) \Bigr] \\
{} \\
\beta_{\tilde{g}_{q^{M}}}^{(nc),dif}= \dfrac{d \tilde{g}_{q^{M}}}{dt} =\dfrac{1}{8\,\pi^{2}} \Bigl[13\, \tilde{g}_{q^{M}}^{3}(t) + 13\, \tilde{g}_{q^{M}}(t) g_{q^{M}}^{2}(t)
+ g_{q^{M}}(t) \tilde{g}_{q^{M}}^{2}(t)   
+ 6\, \tilde{g}_{q^{M}}(t) g_{e^{M}}^{2}(t) + 6 \tilde{g}_{q^{M}}(t) \tilde{g}_{e^{M}}^{2}(t)\Bigr]. \end{array}
\right.
\end{eqnarray}
system of six autonomous coupled cubic ordinary differential equations\footnote{In what follows, all couplings are dimensionless although we drop the index ${}_{Ren.}$.}.

These equations are symmetric in the swap between $g_{_X}$ and $\tilde{g}_{_X}$ couplings belonging to the same vertex.  The index $X$ stands for $M,\, e^{M}$ or $ q^{M}$.
Hence, to get the dependence of the noncommutative mirror Yukawa coupling constants on the renormalization energy scale, we need to solve these coupled differential equations numerically, for given sets of initial values.

Furthermore, it is worth pointing out here that for the special case where $g_{_X}(t) = \tilde{g}_{_X}(t)$, our noncommutative system of the differential equations becomes :
\begin{equation}\label{Beta-function-equations-for-g=gtilde}
\left\{\begin{array}{l}
\beta_{g_{_M}}^{(nc),equ}= \beta_{\tilde{g}_{_M}}^{(nc),equ}= \dfrac{d g_{_M}}{dt} =\dfrac{1}{16\,\pi^{2}} \left[ 17 \, g_{_M}^{3}(t) +  g_{_M}(t) g_{e^{M}}^{2}(t) \right], \\
{}\\
\beta_{g_{e^{M}}}^{(nc),equ}= \beta_{\tilde{g}_{e^{M}}}^{(nc),equ}= \dfrac{d g_{e^{M}}}{dt} =\dfrac{1}{32\,\pi^{2}} \left[ 58 \, g_{e^{M}}^{3}(t) + 3 \, g_{e^{M}}(t)  g_{{_M}}^{2}(t) + 96\, g_{e^{M}}(t) g_{q^{M}}^{2}(t) \right], \\
{}\\
\beta_{g_{q^{M}}}^{(nc),equ}= \beta_{\tilde{g}_{q^{M}}}^{(nc),equ}= \dfrac{d g_{q^{M}}}{dt} =\dfrac{1}{4\,\pi^{2}} \left[13.5\, g_{q^{M}}^{3}(t) + 6\, g_{q^{M}}(t) g_{e^{M}}^{2}(t) \right].
\end{array}
\right.
\end{equation}
A similar computation in the commutative case\footnote{The Feynman diagrams for the commutative case are the same as those for the noncommutative one since no new vertex is introduced by the noncommutativity. All that happens in the noncommutative case is that the Yukawa vertex splits into two orthogonal parts, and the propagators get vulcanised to cure the UV/IR mixing.}, enables us to get the following $\beta$ functions for the corresponding commutative mirror Yukawa couplings ${g_{X}^{(c)}}$ \footnote{We fall back into the Le-Hung commutative result \cite{EWRH-Hung-2014} if we add the Feynman Diagram depicted in the right side of Figure 1 in \cite{EWRH-Hung-2014}, and if in the scalar self energy we consider only one diagram with a mirror leptonic loop for the case of $\beta_{g_{e^M}}$ and a mirror quark loop for the case of $\beta_{g_{q^M}}$, hence inducing the decoupling of the mirror quark's Yukawa $\beta$ function as revealed by Equation (23) in \cite{EWRH-Hung-2014}. }:
\begin{equation} \label{Beta-function-equations-com}
\left\{
    \begin{array}{l}
    \beta_{g_{_M}^{(c)}}= \dfrac{d g_{_M}^{(c)}}{dt} = \dfrac{1}{2} \times \dfrac{1}{16 \,\pi^{2}} \left\{ 19 \, \left[g_{_M}^{(c)}(t)\right]^3 +   g_{_M}^{(c)}(t) \left[g_{e^{M}}^{(c)}(t)\right]^{2} \right\},      \\ {}\\
    \beta_{g_{e^{M}}^{(c)}}= \dfrac{d g_{e^{M}}^{(c)}}{dt} = \dfrac{1}{2} \times \dfrac{1}{ 32\,\pi^{2}} \left\{ 62 \, \left[g_{e^{M}}^{(c)}(t)\right]^{3} + 3 \, g_{e^{M}}^{(c)}(t)  \left[g_{{_M}}^{(c)}(t)\right]^{2} + 96\, g_{e^{M}}^{(c)}(t) \left[g_{q^{M}}^{(c)}(t)\right]^{2}\right\},   \\ {}\\
    \beta_{g_{q^{M}}^{(c)}}= \dfrac{d g_{q^{M}}^{(c)}}{dt} = \dfrac{1}{2} \times \dfrac{1}{ 4\,\pi^{2}} \left\{ 14\, \left[g_{q^{M}}^{(c)}(t)\right]^{3} + 6\, g_{q^{M}}^{(c)}(t) \left[g_{e^{M}}^{(c)}(t)\right]^{2}\right\}. 
    \end{array}
\right.
\end{equation}
These results lead us to the following remarks:
\begin{enumerate}
\item There are six independent noncommutative mirror Yukawa couplings that are coordinated through the $\beta_{g_{_X}}^{(nc),dif}$ functions. This number of independent couplings reduces to three for the special case of $g_{_X}(t) = \tilde{g}_{_X}(t)$ as in the commutative case.
\item The special case where $g_{_X}(t) = \tilde{g}_{_X}(t)$ does not correspond to the commutative case since it always carries information about noncommutative spacetime. This difference seems natural since it can be partly attributed to the fact that even for $g_{_X}(t) = \tilde{g}_{_X}(t)$ the dependence of the noncommutative vertex coupling~(\ref{yukawa-vertex}) on the small noncommutative finite parameter $\theta$ persists and is given by:
\begin{equation} \label{gx-com-noncom}
    g_{_X}\,e^{\frac{i}{2}p\prime \tilde{p}} + \tilde{g}_{_X}\,e^{-\frac{i}{2}p\prime \tilde{p}} \to 2\, g_{_X}\, \cos \left( \frac{1}{2}p^{\prime}_\mu \theta^{\mu\nu} p_\nu \right).
\end{equation}
Besides the fact that in the loop radiative calculations of interest, the involved vulcanised propagators (\ref{vulcanised-scalar-propagator}), (\ref{vulcanised-spinor-propagator}) contain the noncommutative parameters, which can give rise to slightly modified $\beta$ functions compared to the commutative case.
    \item The commutative system of equations (\ref{Beta-function-equations-com})  has coefficients that are roughly half of the corresponding ones in the noncommutative system (\ref{Beta-function-equations-for-g=gtilde}) i.e. for $g_{_X}(t) = \tilde{g}_{_X}(t)$, which means that : 
\begin{equation} \label{equ-57}
    \beta_{g_{_X}^{(c)}} = \dfrac{dg_{_X}^{(c)}(t^\prime)}{dt^\prime} \simeq \dfrac{1}{2} \beta_{g_{_X}^{(nc),equ}} = \dfrac{1}{2} \dfrac{dg_{_X}^{(nc),equ}(t)}{dt}.
\end{equation}
This property can be easily understood by inspecting the formula (\ref{beta-counter-terms}) of the $\beta$ function, and simultaneously by taking into account the substitution formulas at the end of the previous section 
and the one loop Feynman diagrams of Figures \ref{fig:Feynman-diagrams-leptonic-corrections}, \ref{fig:Feynman-diagrams-quark-corrections}, and \ref{fig:Feynman-diagrams-neutrino-corrections} presented in the Appendix.
\begin{enumerate}
\item In fact, out of respect for the loop order substitution relations presented at the end of the previous section, when passing from commutative to noncommutative functions $\beta$ for $g_{_X}(t) = \tilde{g}_{_X}(t)$, we must perform the following substitutions:
    \begin{equation} \label{special-case-substitutions}
        \left\{
        \begin{array}{ll}
            \left[g_{_X}^{(c)}\right]^2 \to \left[g_{_X}^{(nc),equ}\right]^2, &\qquad \text{ in the vertex counter term},  \\
            {} \\
            \left[g_{_X}^{(c)}\right]^2 \to 2\, \left[g_{_X}^{(nc),equ}\right]^2, &\qquad \text{ in the propagator counter terms}.
        \end{array}
        \right.
    \end{equation}
    \item Next, looking at the one loop Feynman diagrams in Figures \ref{fig:Feynman-diagrams-leptonic-corrections}, \ref{fig:Feynman-diagrams-quark-corrections} and \ref{fig:Feynman-diagrams-neutrino-corrections} in the Appendix, we notice that for each process of interest, there is only one vertex  Feynman diagram for about ten self energy Feynman diagrams involved. Thus, in our expressions for the $\beta$ functions, the propagators that contribute a factor of two outnumber the vertices that contribute only a factor of one. This roughly translates to a factor of two of the noncommutative $\beta$ functions relative to the corresponding commutative functions, which appears in the relation~(\ref{equ-57}).
\end{enumerate}
\item Moreover, by setting $\theta =0$ in the tree level noncommutative Yukawa vertex relationship (\ref{yukawa-vertex}), we end up with the following correspondence rules between the commutative and noncommutative mirror Yukawa couplings \footnote{Here, we must pay attention to the fact that $g_{_X}^{(c)} = g_{_{X,c}} +\tilde{g}_{_{X,c}}$, while $g_{_X}^{(c)} \neq g_{_{X}} +\tilde{g}_{_{X}}$ since the former are commutative quantities while the later are noncommutative ones. Hence, we will just compare the commutative couplings $g_{_X}^{(c)}(t) $ evolution in term of $t$, with the corresponding noncommutative $g_{_{X}} +\tilde{g}_{_{X}}$ ones (i.e $g_{_X}^{(c)} \to g_{_{X}} +\tilde{g}_{_{X}}$), in order to identify discrepancies between them. 
Also, one should not confuse the factors 2 that appear in (\ref{special-case-substitutions}) and (\ref{gxc-gxnc-crude-correspondence}). The first originates from a loop level computation while the second comes from a tree level formula.
}:
\begin{equation}\label{gxc-gxnc-crude-correspondence}
\left\{\begin{array}{ll}
 g_{_X}^{(c)}= g_{_{X,c}} +\tilde{g}_{_{X,c}} \to g_{_X} +  \tilde{g}_{_X},  & \qquad \text{for}\quad g_{_X}(t) \neq \tilde{g}_{_X}(t).\\
{}\\
 g_{_X}^{(c)}= 2\, g_{_{X,c}} \to 2\,g_{_X},  & \qquad \text{for}\quad g_{_X}(t) = \tilde{g}_{_X}(t).
\end{array}
\right.
\end{equation}
Hence, to look for the discrepancy between the noncommutative and the commutative mirror Yukawa couplings as a function of the renormalization scale, we will compare the combined noncommutative Yukawa mirror fine structure $\alpha_X^{(nc)} = \dfrac{\left[g_{_X} +\tilde{g}_{_X} \right]^2}{4\pi}$ to the commutative one: $\alpha_X^{(c)} = \dfrac{\left[g_{_X}^{(c)}\right]^2}{4\pi}$.
\end{enumerate}
\section{Numerical solutions of the \texorpdfstring{$\beta$}{} functions systems}
\label{section-6}
\begin{figure}[h!]
     \centering
     \begin{subfigure}[b]{0.46\textwidth}
         \centering
         \includegraphics[width=\textwidth]{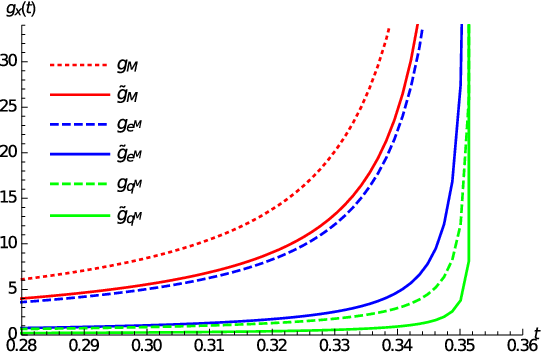}
         \caption{$g_{_X}^{(nc),dif}$ for Set 1 of initial values.}
         \label{g-couplings-all}
     \end{subfigure}
     \hfill
     \begin{subfigure}[b]{0.46\textwidth}
         \centering
         \includegraphics[width=\textwidth]{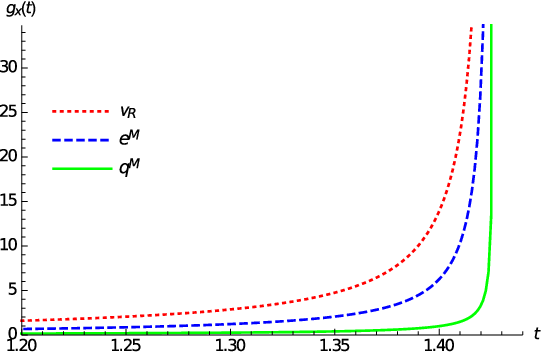}
         \caption{$g_{_X}^{(nc),equ}$ for Set 2 of initial values. }
         \label{alpha-couplings-all}
     \end{subfigure}
 \caption{\small Evolution of the mirror Yukawa couplings as a function of the renormalization scale $t$.}
        \label{mirror-yukawa-couplings}
\end{figure}

Solving numerically the previously obtained $\beta$ systems of equations \footnote{The numerical calculations and the draw of the corresponding plots are performed through Mathematica \cite{mathematica-2014}} enables us to display the dependence of the mirror Yukawa couplings in terms of the $t$ renormalization scale. Figure~\ref{mirror-yukawa-couplings} shows this dependence, in the noncommutative case both for $g_{_X}^{(nc),dif}$ and for $g_{_X}^{(nc),equ}$~\footnote{Where $g_{_X}^{(nc),dif}$ stands for $g_{_X}$ or $\tilde{g}_{_X}$  with $g_{_X} \neq \tilde{g}_{_X}$, and $g_{_X}^{(nc),equ}$ also for $g_{_X}$ or $\tilde{g}_{_X}$ but with $g_{_X} = \tilde{g}_{_X}$.}, respectively for Set 1 and Set 2 of values listed in Table \ref{tab:used-initial-values-sets}, and which are related to the initial mirror Yukawa coupling values through the following relations:
\begin{equation}
\left\{
    \begin{array}{lll}
    g_{_M}^{(nc),\,dif}(0) = x \,,  & \qquad g_{e^M}^{(nc),\,dif}(0) = y\,,   &  \qquad g_{q^M}^{(nc),\,dif}(0) = z\,, \\
    {} \\
    \tilde{g}_{_M}^{(nc),\,dif}(0) = \tilde{x}\,,     & \qquad \tilde{g}_{e^M}^{(nc),\,dif}(0) = \tilde{y}\,, & \qquad \tilde{g}_{q^M}^{(nc),\,dif}(0) = \tilde{z}\,,
    \end{array}
\right.
\end{equation}

\begin{equation}
    \left\{
    \begin{array}{ll}
     g_{_M}^{(nc),\,equ}(0) =\tilde{g}_{_M}^{(nc),\,equ}(0)= (x + \tilde{x})/2   \,, &
     \qquad g_{e^M}^{(nc),\,equ}(0) =\tilde{g}_{e^M}^{(nc),\,equ}(0)= (y + \tilde{y})/2  \,, \\
     {} \\
     g_{q^M}^{(nc),\,equ}(0) =\tilde{g}_{q^M}^{(nc),\,equ}(0)= (z + \tilde{z})/2 \,, &
    \end{array}
    \right.
\end{equation}

\begin{equation}
    g_{_M}^{(c)}(0) = x + \tilde{x}\,, \qquad g_{e^M}^{(c)}(0) = y + \tilde{y}\,, \qquad g_{q^M}^{(c)}(0) = z + \tilde{z}\,.
\end{equation}

\begin{table}[!h]
    \centering
    \begin{tabular}{ccccccc} \hline
        & $x$ & $\tilde{x}$ & $y$ & $\tilde{y}$ & $z$ & $\tilde{z}$ \\ \hline
     Set 1  & 3.95 & 3.2 & 2.95 & 1.35 & 1.41 & 0.82 \\ \hline
       Set 2  & 1.784 & 1.784 & 1.1 & 1.1 & 0.617 & 0.617 \\ \hline
        Set 3  & 3.9 & 1.6 & 1.26 & 2.2 & 1.25 & 0.8 \\ \hline
           Set 4  & 2.3 & 0.2 & 2.0 & 0.46 & 1.7 & 1.03 \\ \hline
            Set 5  & 2.32 & 0.5 & 1.1 & 0.48 & 1.9 & 1.06 \\ \hline
                Set 6  & 0.92 & 0.25 & 1.02 & 0.35 & 1.12 & 0.45 \\ \hline

    \end{tabular} 
    \caption{Used constant sets to construct the corresponding initial values mirror Yukawa couplings~\protect\footnotemark.}
    \label{tab:used-initial-values-sets}
\end{table}
 \footnotetext{In choosing these sets of constants, we ensured that the initial values of all commutative Yukawa couplings are greater than the Yukawa top quark coupling which approaches unity. Moreover Set 4 and Set 5 correspond, in the commutative case, respectively to the initial values taken by Le-Hung in \cite{EWRH-Hung-2014} and \cite{EWRH-Hung-2016}.}

\begin{figure}[h!]
     \centering
     \begin{subfigure}[b]{0.46\textwidth}
         \centering
         \includegraphics[width=\textwidth]{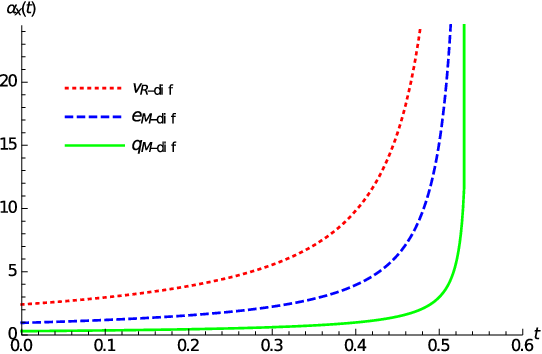}
         \caption{Mirror noncommutative $\alpha_{_X}^{(nc)}(t)$ for \\ $g_{_X} \neq \tilde{g}_{_X}$ }
         \label{fig: alphax g1 neq g1tilde}
     \end{subfigure}
     \hfill
     \begin{subfigure}[b]{0.46\textwidth}
         \centering
         \includegraphics[width=\textwidth]{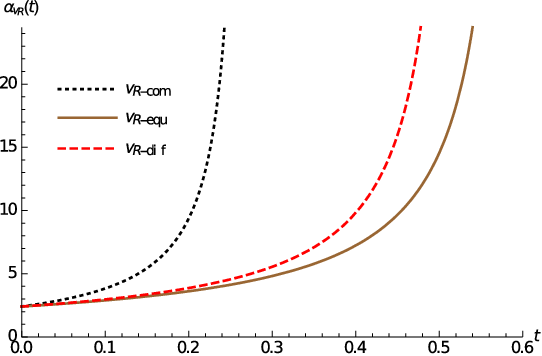}
         \caption{Mirror neutrinos' commutative and noncommutative $\alpha_{\nu_{_R}}(t)$ }
         \label{fig:rhox g1 neq g1tilde}
     \end{subfigure}
     \begin{subfigure}[b]{0.46\textwidth}
         \centering
         \includegraphics[width=\textwidth]{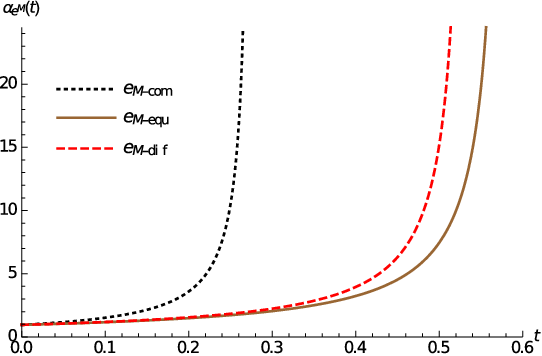}
         \caption{Mirror down lepton's commutative and noncommutative $\alpha_{e^{\scriptscriptstyle M}}(t)$   }
         \label{fig:alphax g1 eq g1tilde}
     \end{subfigure}
     \hfill
     \begin{subfigure}[b]{0.46\textwidth}
         \centering
         \includegraphics[width=\textwidth]{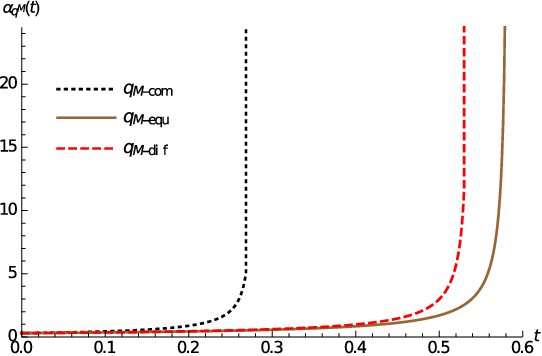}
         \caption{Mirror quark's commutative and noncommutative $\alpha_{q^{\scriptscriptstyle M}}(t)$ }
         \label{fig:rhox g1 eq g1tilde}
     \end{subfigure}
        \caption{\small Evolution of the mirror Yukawa fine structure constants in terms of the renormalization scale $t$ for Set 3 of initial values.}
        \label{fig: alphax-and-rhox}
\end{figure}

Moreover, to smooth the comparison between the commutative and the noncommutative behaviors, it's more convenient to deal with the mirror fine structure constants $\alpha_X^{(nc)}(t)$ and $\alpha_X^{(c)}(t)$, for which the previous initial mirror Yukawa couplings values relationship to the constant sets of Table \ref{tab:used-initial-values-sets} have been set to ensure that~:
\vspace{-0.1cm}
\begin{equation}
    \alpha_X^{(nc), dif}(0) = \alpha_X^{(nc),equ}(0) = \alpha_X^{(c)}(0),
\end{equation}
as shown in Figures \ref{fig:rhox g1 neq g1tilde}, \ref{fig:alphax g1 eq g1tilde} and \ref{fig:rhox g1 eq g1tilde} for the plot's starting point $t=0$.
\begin{table}[!h]
\small
    \centering
    \begin{tabularx}{\textwidth}{Xccccccccccccc} \hline
        & \multicolumn{6}{c}{ All involved loops and flavor families} & \multicolumn{6}{c}{ Only one special loop is considered in } \\ 
                            &\multicolumn{6}{c}{are considered into the scalar propagator} & \multicolumn{6}{c}{the scalar propagator as in Le-Hung \cite{EWRH-Hung-2014} }\\ \hline
             Initial values      & Set 1 & Set 2 & Set 3 & Set 4 & Set 5 & Set 6  & Set 1 & Set 2 & Set 3  & Set 4 & Set 5 & Set 6   \\ \hline 
        $t_{_P}^{(nc),dif}$ & 0.35 & 1.43 & 0.53 & 0.50  & 0.55  & 1.50 & 0.66 & 2.68 & 1.03  & 3.68 & 3.44 & 11.55 \\ \hline 
        $t_{_P}^{(nc),equ}$  & 0.36 & 1.43 & 0.58 & 0.58 & 0.60  & 1.78 & 0.67 & 2.68 & 1.13  & 4.24 & 3.60 & 12.81 \\ \hline
        $t_{_P}^{(c)}$       & 0.16 & 0.64 & 0.27 & 0.28  & 0.29 & 0.87 & 0.27 & 1.10 & 0.46  & 1.77  & 1.50 & 5.34 \\ \hline
    \end{tabularx}
    \caption{\small Landau pole positions for different sets of initial values of mirror Yukawa couplings.}
     \label{tab:temps}
\end{table}

Our results presented in Figures \ref{mirror-yukawa-couplings} and \ref{fig: alphax-and-rhox}, and Table \ref{tab:temps} highlight the following remarks :
\begin{enumerate}

\item[i)] The noncommutative mirror Yukawa couplings preserve the behavior of the commutative ones pointed in \cite{EWRH-Hung-2014} i.e. a plateau for low $t$ values, then a Landau pole at some higher $t_{_P}$ scale, as shown in Figure~\ref{mirror-yukawa-couplings} for the mirror Yukawa couplings and in Figure~\ref{fig: alphax-and-rhox} in terms of the corresponding mirror Yukawa fine structure constants. 

\item[ii)] The position of the Landau poles in the noncommutative and commutative calculation approaches depends on the chosen set of initial values as shown in Table \ref{tab:temps}. In particular, columns 7, 11, and 12 of this table show that we can set the initial values of the mirror Yukawa coupling so that the noncommutative Landau poles (namely $t_{_P}^{(nc),dif} = 1.50$ and $t_{_P}^{(nc),equ} = 1.78$ for Set 6 of initial values) be the same as the commutative one (that is $t_{_P}^{(c)} = 1.50$ for Set 5 of initial values, and $t_{_P}^{(c)} = 1.77$ for Set 4 of initial values). The above Landau pole values are of special interest because they would correspond to an energy scale in the TeV domain, besides they refer to the values pointed by Le-Hung in \cite{EWRH-Hung-2014} and \cite{EWRH-Hung-2016}. 

\item[iii)] Comparing the right side part to the left side part of Table \ref{tab:temps} shows that for the same set of initial values, the fewer loops taken into account in the scalar propagator the higher the values of the Landau poles.

\item[iv)]  The swap symmetry between $g_{_X}$ and $\tilde{g}_{_X}$ couplings in the differential equations manifests itself in the corresponding solutions through the three following ways :
    \begin{enumerate}
        \item In Figure \ref{g-couplings-all} belonging to Set 1 of initial values, the plots for $g_{_X}$ and $\tilde{g}_{_X}$ are interchanged when we swap the corresponding initial values.
        \item When we adopt Set 2 of initial values in Equations~(\ref{Beta-function-equations-non-com}) and  (\ref{Beta-function-equations-for-g=gtilde}), their corresponding solutions become superimposed and give the plots drawn in Figure~\ref{alpha-couplings-all}.
        \item For Set 2 of initial values, the Landau poles $t_{_P}^{(nc),dif}$ and $t_{_P}^{(nc),equ}$ arising respectively from Equations (\ref{Beta-function-equations-non-com}) and  (\ref{Beta-function-equations-for-g=gtilde}) become identical as indicated by the corresponding columns in Table \ref{tab:temps}.
    \end{enumerate}
    
\item[v)] Table \ref{tab:temps} and Figures \ref{fig:rhox g1 neq g1tilde}, \ref{fig:alphax g1 eq g1tilde}, and \ref{fig:rhox g1 eq g1tilde} show that the noncommutative Landau poles' positions are roughly twice the corresponding commutative ones for the same initial values. This can be seen as follows:

    For simplicity's sake, we consider the right hand side of the noncommutative equation system (\ref{Beta-function-equations-for-g=gtilde}) and take into account the relationship (\ref{gxc-gxnc-crude-correspondence}) that connect the commutative and the noncommutative couplings, i.e. by plugging $\dfrac{g_{_X}^{(c)}(t^\prime)}{2}$ for each $g_{_X}(t)$. We end up with an additional $\dfrac{1}{8}$ global factor on the right hand side due to cubic terms, and a factor $\dfrac{1}{2}$ on the left hand side. Which, for the first equation in (\ref{Beta-function-equations-for-g=gtilde}) leads to:
    \begin{equation} \label{equ-1}
    \dfrac{dg_{_M}^{(nc),equ}(t)}{dt} \to \dfrac{1}{2} \dfrac{dg_{_M}^{(c)}(t^\prime)}{dt}
    =\dfrac{1}{16\,\pi^{2}} \left[ \dfrac{17}{8} \, \left\{ g_{_M}^{(c)} \right\}^{3}(t^\prime) + \dfrac{1}{8} \,g_{_M}^{(c)}(t^\prime) \left\{ g_{e^{M}}^{(c)} \right\}^{2}(t^\prime) \right].
    \end{equation}
    Or equivalently:
    \begin{equation} \label{equ-2}
    \dfrac{dg_{_M}^{(c)}(t^\prime)}{d\left(t/2\right)}
    =\dfrac{1}{16\,\pi^{2}} \left[ \dfrac{17}{2} \, \left\{ g_{_M}^{(c)} \right\}^{3}(t^\prime) + \dfrac{1}{2} \, g_{_M}^{(c)}(t^\prime) \left\{ g_{e^{M}}^{(c)} \right\}^{2}(t^\prime) \right].
    \end{equation}
    Which must be brought closer to the right hand side of the first equation in (\ref{Beta-function-equations-com}), i.e.
   \begin{equation} \label{equ-3}
    \dfrac{dg_{_M}^{(c)}(t^\prime)}{dt^\prime} =  \dfrac{1}{16 \,\pi^{2}} \left\{ \dfrac{19}{2} \, \left[g_{_M}^{(c)}(t^\prime)\right]^3 + \dfrac{1}{2} \,  g_{_M}^{(c)}(t^\prime) \left[g_{e^{M}}^{(c)}(t^\prime)\right]^{2} \right\} \,.
   \end{equation}
    Hence, if in Equation (\ref{equ-2}) we set:
    \begin{equation} \label{com-noncom-scales}
    t = 2\,t^\prime ,
    \end{equation}
    we end up with an equation that has a similar form to Equation (\ref{equ-3}) with the coefficient $19$ replaced by the coefficient $17$, and with $t$ linked to the noncommutative case and $t^\prime$ to the commutative case. So, if we model a simple commutative Landau pole by:
    \begin{equation}
    \lim_{t^\prime \to t_{_P}^{(c)}} \frac{A}{B - Ct^\prime } \to +\infty, \quad \text{with}\,\, A>0,\,\, B>0, \,\,\text{and}\,\, C>0,
    \end{equation}
    that is :
    \begin{equation}
        t_{_P}^{(c)} = \frac{B}{C}\,.
    \end{equation}
    In addition, we weaken constraint (\ref{com-noncom-scales}) so that the equality it contains is no longer exact but rather approximate. The corresponding noncommutative Landau pole will then manifest itself at :
    \begin{equation}
    t_{_P}^{(nc)} \approx 2\, t_{_P}^{(c)}\,,
    \end{equation}
    in complete agreement with the results cited above. 

\item[vi)] Table \ref{tab:temps} and Figures \ref{fig:rhox g1 neq g1tilde}, \ref{fig:alphax g1 eq g1tilde}, and \ref{fig:rhox g1 eq g1tilde} show that for each of the sets 1, 3, 4, 5, and 6 of initial values that fulfill for $t=0$ the relation $g_{_X}^{(nc),\,dif}(0) \neq \tilde{g}_{_X}^{(nc),\,dif}(0)$, the noncommutative Landau poles $t_{_P}^{(nc),dif}$ and $t_{_P}^{(nc),equ}$ belonging respectively to Equations (\ref{Beta-function-equations-non-com}) and~(\ref{Beta-function-equations-for-g=gtilde}) have distinct values as expected, but with $t_{_P}^{(nc),equ}$ always greater than~$t_{_P}^{(nc),dif}$, while for Set 2 of initial values for which at $t=0$ we have $g_{_X}^{(nc),\,dif}(0) = \tilde{g}_{_X}^{(nc),\,dif}(0)$, the pole positions $t_{_P}^{(nc),dif}$ and $t_{_P}^{(nc),equ}$ are the same.

To understand this property, we focus on the Landau pole region for which $t$ is in the interval~$]t_{_0}, t_{_P}[ $ with $t_{_0}$ having a value lower than but close to $t_{_P}$. In this region, all the Yukawa coupling functions  $g_{_X}(t) $ and $\tilde{g}_{_X}(t)$ carry large values. 
Let us therefore consider two functions $\delta(t)$ and $\eta(t)$ that carry large values and have behavior that mimics the mirror Yukawa couplings in this region. Furthermore, keeping in mind the above swap symmetry, we can write~:
\begin{equation}
    \left.
    \begin{array}{l}
       g_{_X}^{(nc),dif}(t) = \delta(t) + \epsilon \, \,,  \\
       {} \\
       \tilde{g}_{_X}^{(nc),dif}(t) = \delta(t) - \epsilon  \,\,,  \\
       {} \\
       g_{_X}^{(nc),equ}(t) = \tilde{g}_{_X}^{(nc),equ}(t) = \eta(t)\,\,,
    \end{array}
    \right\} \quad \text{for} \,\, t \in \,\, ]t_{_0}, t_{_P}[\,,
\end{equation}
regardless of $X$ being $M$, $e_M$, or $q_M$, and with $\epsilon$ being an infinitesimal positive parameter.

In addition, using the first equation in the equations system~(\ref{Beta-function-equations-non-com}) and the first equation in the equations system~(\ref{Beta-function-equations-for-g=gtilde}), we obtain the following~: 
\begin{equation}
    \left\{
    \begin{array}{l}
     (\ref{Beta-function-equations-non-com}) \quad \to  \quad \dfrac{d\delta(t)}{dt} = \dfrac{1}{8 \pi^2}
     \left[   4[\delta(t) + \epsilon]^3 + 4 [\delta(t) + \epsilon] [\delta(t) - \epsilon]^2 +  [\delta(t) + \epsilon]^2[\delta(t) - \epsilon]
     \right]  \,\,, \\
    {} \\
    (\ref{Beta-function-equations-for-g=gtilde}) \quad \to \quad \dfrac{d\eta(t)}{dt} =  \dfrac{9}{8 \pi^2}\,\eta^3(t)  \,\,.
     \end{array}
     \right.
\end{equation}

So, to first order in the infinitesimal parameter $\epsilon$, the above equations take the form~:
\begin{equation}
    \left\{
    \begin{array}{l}
         \dfrac{d\delta(t)}{dt} = (1+\epsilon)\, \dfrac{9}{8 \pi^2} \, \delta^3(t) \,\,,  \\
         {} \\
         \dfrac{d\eta(t)}{dt} =  \dfrac{9}{8 \pi^2}\, \eta^3(t) \,\,.
    \end{array}
    \right.
\end{equation}
Thus, for a given $t$ in \,\, $]t_{_0}, t_{_P}[$, the slope $\dfrac{d\delta(t)}{dt}$ of the function $\delta(t)$ is slightly greater than the slope $\dfrac{d\eta(t)}{dt}$ of the function $\eta(t)$. Therefore, assuming $\delta(t_0) = \eta(t_0)$ we get that $\delta(t)$ will reach its asymptote i.e.~$t_{_P}^{(nc),dif}$ sooner than $\eta(t)$ does i.e. $t_{_P}^{(nc),equ}$.

We should then have~:
\begin{equation}
    t_{_P}^{(nc),equ} \, \geq  \, t_{_P}^{(nc),dif},
\end{equation}
which is what we obtain numerically. 
\end{enumerate}
\vspace{-0.44cm}
\section{Concluding remarks}
\label{conclusion}
In this work, we have first used the Moyal product to construct a noncommutative Euclidean mirror Yukawa Lagrangian within the electroweak-scale right handed neutrinos model.

Secondly, we established the Slavnov-Taylor identities for the noncommutative universal mirror Yukawa couplings. We utilized them to select from several possibilities the processes for which we had to calculate the one loop order radiative corrections involved in determining the counter terms for the noncommutative mirror Yukawa $\beta$ functions. Through this calculation, the noncommutative vulcanised scalar and spinor propagators have provided valuable support to avoid the mixing between the ultraviolet and infrared momenta.

The obtained counter term ultraviolet divergent parts results only from planar integrals and thus have the same divergent structure than the commutative case but with different factors. Moreover, they do not depend explicitly on the deformation parameter $\theta$, nor on the vulcanised parameters $a$ and $b$. These parameters only affect the finite part for which both planar and non-planar integrals contribute.
This means that the commutative limit cannot be obtained simply by making the noncommutative parameters tend to zero in which case this leads us to expressions which no longer make sense.

However, analyzing the expressions of these counter terms in parallel with their implication in the expression of the $\beta$ function allows us to state a simple relationship that~enables a direct link between the commutative and the noncommutative one loop $\beta$ function~expressions in terms of occurring couplings. These correspondence rules act differently for propagators and vertices and thus reflect the fact that there can be no strict proportionality between the considered commutative and noncommutative $\beta$ functions, but rather a possible approximate proportionality depending on the relative weights between the involved counter~terms.

Our procedure results in a system of six coupled nonlinear first order differential equations, that we solved numerically. The obtained numerical solutions allowed us, to state that at the one loop level within the EW$\nu_R^M$ and in terms of renormalization scale, the noncommutative geometry preserves the behavior of a plateau followed by a Landau pole for the mirror Yukawa couplings as is the case in the commutative formulation, with the notable difference that the position of the noncommutative Landau pole is roughly twice of the commutative one. This would indicate that in noncommutative spacetime, new phenomena can be triggered by a phase transition at higher energy scales than would be expected by the commutative case. 

Moreover, the exact behavior of the obtained solutions and the position of the corresponding Landau pole depends on the choice of the initial values that may be related to low energy masses of the mirror fermions. 
Therefore, it seems interesting to probe the inverse problem in a multidimensional coupling space to shed more light on the flow between high and low energy scales. This is left for future work.

Finally, exploring the effects of the $\lambda_i$ couplings occurring in the scalar potential $V(\varphi_s, \Phi_2, \Phi_{2M}, X)$, on the running of the noncommutative mirror Yukawa couplings $g_X$ and~$\tilde{g}_X$, is a natural extension of this work since the Yukawa interaction is involved in the one loop Feynman diagrams corresponding to this potential. This is currently under investigation and will be discussed separately in the future.

\appendix
\makeatletter
\def\@seccntformat#1{\csname Pref@#1\endcsname \csname the#1\endcsname\quad}
\def\Pref@section{Appendix~}
\makeatother
\section{Radiative corrections leading to the counter terms (\ref{equ-counter-term-gem-gemtilde})-(\ref{equ-counter-term-chi0-nuR})}
Applying noncommutative vulcanised Feynman rules to the Feynman diagrams depicted in Figures \ref{fig:Feynman-diagrams-leptonic-corrections}, \ref{fig:Feynman-diagrams-quark-corrections} and~\ref{fig:Feynman-diagrams-neutrino-corrections} belonging respectively to the processes 
{\footnotesize $ e_{R}^{M}(p) + \overline{e}_{L}^{M}(p^\prime) \to \phi_{2M}^{0}(q)$}, 
 {\footnotesize $u_{R}^{M}(p) + \overline{u}_{L}^{M}(p^\prime) \to \phi_{2M}^{0}(q)$}
 , and {\footnotesize $ \nu_{e_R}(p) + \nu_{e_R}^{T}(p^\prime) \to \chi^{0}(q)$}, leads to the following radiative corrections:
 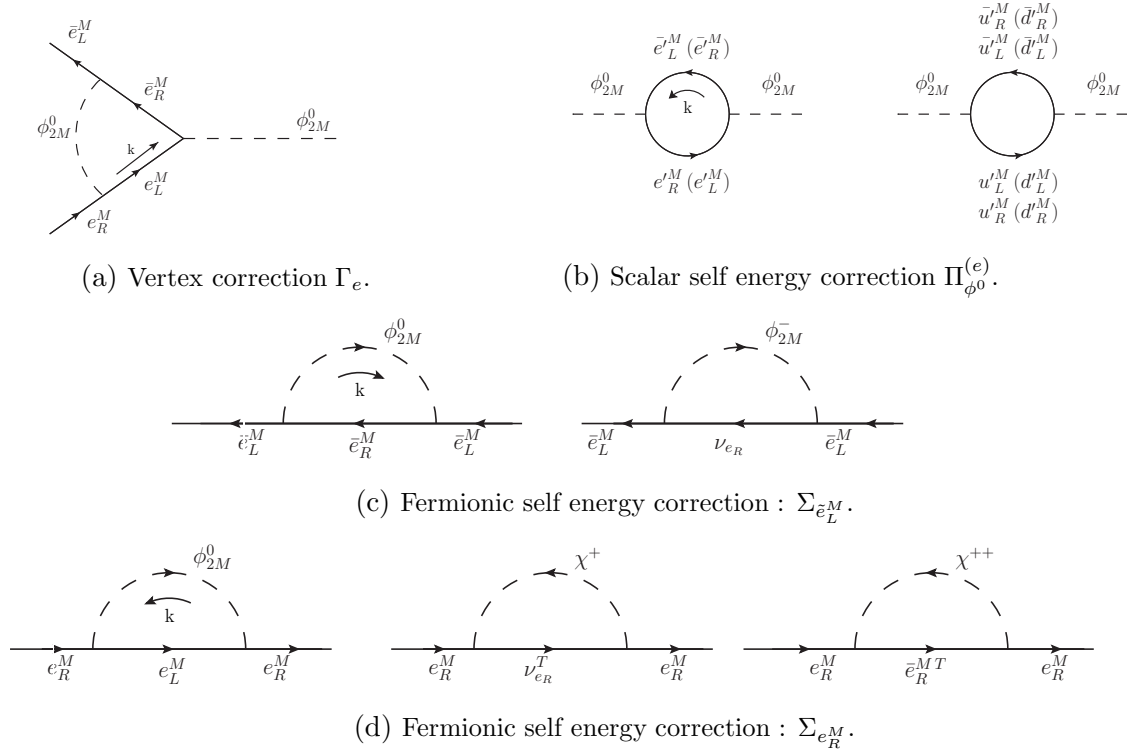
\begin{figure}[h!]
     \centering
     \begin{subfigure}[b]{0.3\textwidth}
         \centering
\scalebox{.45}{ 
\fcolorbox{white}{white}{

  \begin{picture}(352,282) (133,-123)
    \SetWidth{1.0}
    \SetColor{Black}
    \Line[arrow,arrowpos=0.65,arrowlength=5,arrowwidth=2,arrowinset=0.2,clock](144,-122)(256,-42)
    \Line[arrow,arrowpos=0.20,arrowlength=5,arrowwidth=2,arrowinset=0.2,clock](144,-122)(256,-42)
    \Line[arrow,arrowpos=0.80,arrowlength=5,arrowwidth=2,arrowinset=0.2](256,-42)(144,38)
    \Line[arrow,arrowpos=0.35,arrowlength=5,arrowwidth=2,arrowinset=0.2](256,-42)(144,38)
    \Line[dash,dashsize=10](256,-42)(384,-42)
    \Arc[dash,dashsize=10](232,-40)(66,135,228)
    \Text(160,38)[lb]{\Large{\Black{$ \bar{e}_L^M$}}}
    \Text(224,-10)[lb]{\Large{\Black{$ \bar{e}_R^M$}}}
    \Text(352,-36)[lb]{\Large{\Black{$\phi_{2M}^{0}$}}}
    \Text(134,-42)[lb]{\Large{\Black{$\phi_{2M}^{0}$}}}
    \Text(224,-90)[lb]{\Large{\Black{$ {e}_L^M $}}}
            \Line[arrow,arrowpos=0.90,arrowlength=5,arrowwidth=2,arrowinset=0.2,clock](200,-72)(235,-45)
              \Text(210,-55)[lb]{k}
    \Text(176,-122)[lb]{\Large{\Black{$e_R^M$}}}
  \end{picture}
}}
         \caption{\footnotesize Vertex correction  $\Gamma_e$.}
         \label{fig:Feynman-Diagrams-vertex-leptons}
     \end{subfigure}  \hspace{-1.0cm}
     \begin{subfigure}[b]{0.6\textwidth}
         \centering
\scalebox{.45}{ 
\fcolorbox{white}{white}{

  \begin{picture}(308,166) (143,-159)
    \SetWidth{1.0}
    \SetColor{Black}
    \Text(416,-42)[lb]{\Large{\Black{$\phi_{2M}^{0}$}}}
    \Line[dash,dashsize=10](256,-58)(317,-58)
    \Arc[arrow,arrowpos=0.45,arrowlength=5,arrowwidth=2,arrowinset=0.2](352,-58)(35,117,477)
        \Arc[arrow,arrowpos=0.95,arrowlength=5,arrowwidth=3,arrowinset=0.2](352,-58)(20,45,135)
    \Arc[arrow,arrowpos=0.92,arrowlength=5,arrowwidth=2,arrowinset=0.2](352,-58)(35,117,477)
    \Line[dash,dashsize=10](387,-58)(448,-58)
    \Text(272,-42)[lb]{\Large{\Black{$\phi_{2M}^{0}$}}}
    \Text(350,-57)[lb]{\large k}
    \Text(325,-12)[lb]{\Large{\Black{$\bar{e^\prime}_L^M \, (\bar{e^\prime}_R^M)$}}}
    \Text(325,-125)[lb]{\Large{\Black{${e^\prime}_R^M \, ({e^\prime}_L^M)$}}}
    \end{picture} \hspace{-1.5cm}
  
    \begin{picture}(308,166) (143,-159)
    \SetWidth{1.0}
    \SetColor{Black}
    \Text(416,-42)[lb]{\Large{\Black{$\phi_{2M}^{0}$}}}
    \Line[dash,dashsize=10](256,-58)(317,-58)
    \Arc[arrow,arrowpos=0.45,arrowlength=5,arrowwidth=2,arrowinset=0.2](352,-58)(35,117,477)
    \Arc[arrow,arrowpos=0.92,arrowlength=5,arrowwidth=2,arrowinset=0.2](352,-58)(35,117,477)
    \Line[dash,dashsize=10](387,-58)(448,-58)
    \Text(272,-42)[lb]{\Large{\Black{$\phi_{2M}^{0}$}}}
    \Text(325,+13)[lb]{\Large{\Black{$\bar{u^\prime}_R^M \, (\bar{d^\prime}_R^M)$}}}
    \Text(325,-12)[lb]{\Large{\Black{$\bar{u^\prime}_L^M \, (\bar{d^\prime}_L^M)$}}}
    \Text(325,-150)[lb]{\Large{\Black{${u^\prime}_R^M \, ({d^\prime}_R^M)$}}}
    \Text(325,-125)[lb]{\Large{\Black{${u^\prime}_L^M \, ({d^\prime}_L^M)$}}}
  \end{picture}
}}
         \caption{\footnotesize Scalar self energy correction $\Pi_{\phi^0}^{(e)}$.}
         \label{fig:Feynman-Diagrams-scalar-self-energy-leptons}
     \end{subfigure}
     \begin{subfigure}[b]{0.90\textwidth}
         \centering
\scalebox{.60}{ 
\fcolorbox{white}{white}{
\begin{picture}(204,96) (143,-102)
    \SetWidth{1.0}
    \SetColor{Black}
    \Line[arrow,arrowpos=0.13,arrowlength=5,arrowwidth=2,arrowinset=0.2](310,-80)(108,-80)
    \Line[arrow,arrowpos=0.45,arrowlength=5,arrowwidth=2,arrowinset=0.2](304,-80)(90,-80)
    \Line[arrow,arrowpos=0.99,arrowlength=5,arrowwidth=2,arrowinset=0.2](304,-80)(128,-80)
    \Arc[dash,dashsize=10,arrow,arrowpos=0.5,arrowlength=5,arrowwidth=2,arrowinset=0.2,clock](208,-80)(48,-180,-360)
        \Arc[arrow,arrowpos=0.95,arrowlength=5,arrowwidth=2,arrowinset=0.2,clock](208,-80)(32,-245,65)
        \Text(206,-63)[lb]{k}
    \Text(224,-30)[lb]{\large{\Black{$\phi_{2M}^{0}$}}}
    \Text(132,-100)[lb]{\large{\Black{$\bar{e}_L^M$}}}
    \Text(202,-102)[lb]{\large{\Black{$\bar{e}_R^M$}}}
    \Text(268,-100)[lb]{\large{\Black{$\bar{e}_L^M$}}}
  \end{picture} 
 \hspace{1.0cm}
\begin{picture}(204,96) (143,-102)
    \SetWidth{1.0}
    \SetColor{Black}
     \Line[arrow,arrowpos=0.10,arrowlength=5,arrowwidth=2,arrowinset=0.2](310,-80)(108,-80)
    \Line[arrow,arrowpos=0.55,arrowlength=5,arrowwidth=2,arrowinset=0.2](304,-80)(128,-80)
     \Line[arrow,arrowpos=0.95,arrowlength=5,arrowwidth=2,arrowinset=0.2](304,-80)(128,-80)
    \Arc[dash,dashsize=10,arrow,arrowpos=0.5,arrowlength=5,arrowwidth=2,arrowinset=0.2,clock](208,-80)(48,-180,-360)
    \Text(224,-30)[lb]{\large{\Black{$\phi_{2M}^{-}$}}}
    \Text(112,-100)[lb]{\large{\Black{$\bar{e}_L^M$}}}
    \Text(192,-100)[lb]{\large{\Black{$\nu_{e_R}$}}}
    \Text(262,-100)[lb]{\large{\Black{$\bar{e}_L^M$}}}
  \end{picture}
}}
         \caption{\footnotesize Fermionic self energy correction : $\Sigma_{\tilde{e}_L^M}$.}
         \label{fig:Feynman-Diagrams-fermionic-self-energy-leptons-left}
\end{subfigure}
\begin{subfigure}[b]{0.90\textwidth}
\centering
\scalebox{.60}{ 
\fcolorbox{white}{white}{
  \begin{picture}(204,96) (143,-102)
    \SetWidth{1.0}
    \SetColor{Black}
    \Line[arrow,arrowpos=0.15,arrowlength=5,arrowwidth=2,arrowinset=0.2](108,-80)(310,-80)
     \Line[arrow,arrowpos=0.45,arrowlength=5,arrowwidth=2,arrowinset=0.2](128,-80)(304,-80)
    \Line[arrow,arrowpos=0.90,arrowlength=5,arrowwidth=2,arrowinset=0.2](128,-80)(304,-80)
    \Arc[dash,dashsize=10,arrow,arrowpos=0.5,arrowlength=5,arrowwidth=2,arrowinset=0.2,clock](208,-80)(48,-180,-360)
        \Arc[arrow,arrowpos=0.95,arrowlength=5,arrowwidth=2,arrowinset=0.2](208,-80)(32,65,115)
        \Text(206,-63)[lb]{k}
    \Text(224,-30)[lb]{\large{\Black{$\phi_{2M}^{0}$}}}
    \Text(132,-100)[lb]{\large{\Black{$e_R^M$}}}
    \Text(202,-102)[lb]{\large{\Black{${e}_L^M$}}}
    \Text(268,-100)[lb]{\large{\Black{$e_R^M$}}}
  \end{picture}
  \hspace{1.0cm}
  \begin{picture}(204,96) (143,-102)
    \SetWidth{1.0}
    \SetColor{Black}
    \Line[arrow,arrowpos=0.15,arrowlength=5,arrowwidth=2,arrowinset=0.2](108,-80)(310,-80)
    \Line[arrow,arrowpos=0.45,arrowlength=5,arrowwidth=2,arrowinset=0.2](128,-80)(304,-80)
    \Line[arrow,arrowpos=0.90,arrowlength=5,arrowwidth=2,arrowinset=0.2](128,-80)(304,-80)
    \Arc[dash,dashsize=10,arrow,arrowpos=0.5,arrowlength=5,arrowwidth=2,arrowinset=0.2](208,-80)(48,360,180)
    \Text(224,-30)[lb]{\large{\Black{$\chi^+$}}}
    \Text(132,-100)[lb]{\large{\Black{$e_R^M$}}}
    \Text(192,-102)[lb]{\large{\Black{${\nu}_{e_R}^T $}}}
    \Text(278,-100)[lb]{\large{\Black{$e_R^M$}}}
  \end{picture}
  \hspace{1.0cm}
\begin{picture}(204,96) (143,-102)
    \SetWidth{1.0}
    \SetColor{Black}
    \Line[arrow,arrowpos=0.10,arrowlength=5,arrowwidth=2,arrowinset=0.2](108,-80)(310,-80)
     \Line[arrow,arrowpos=0.55,arrowlength=5,arrowwidth=2,arrowinset=0.2](90,-80)(304,-80)
    \Line[arrow,arrowpos=0.90,arrowlength=5,arrowwidth=2,arrowinset=0.2](128,-80)(304,-80)
    \Arc[dash,dashsize=10,arrow,arrowpos=0.5,arrowlength=5,arrowwidth=2,arrowinset=0.2](208,-80)(48,360,180)
    \Text(224,-30)[lb]{\large{\Black{$\chi^{++}$}}}
    \Text(132,-100)[lb]{\large{\Black{$e_R^M$}}}
    \Text(192,-102)[lb]{\large{\Black{$\bar{e}_R^{M\,T}$}}}
    \Text(278,-100)[lb]{\large{\Black{$e_R^M$}}}
  \end{picture} 
}}
         \caption{\footnotesize Fermionic self energy correction : $ \Sigma_{e_R^M}$.}
         \label{fig:Feynman-Diagrams-fermionic-self-energy-leptons-right}
     \end{subfigure}
             \caption{\small One loop radiative corrections to the process {\footnotesize $ e_{R}^{M}(p) + \overline{e}_{L}^{M}(p^\prime) \to \phi_{2M}^{0}(q)$}, where $e$ stands for a specific mirror lepton flavor, $u^\prime$, $d^\prime$, and $e^\prime$ stand each for respectively the three mirror up quark, three mirror down quark, and three mirror lepton flavors.}
        \label{fig:Feynman-diagrams-leptonic-corrections}
\end{figure}
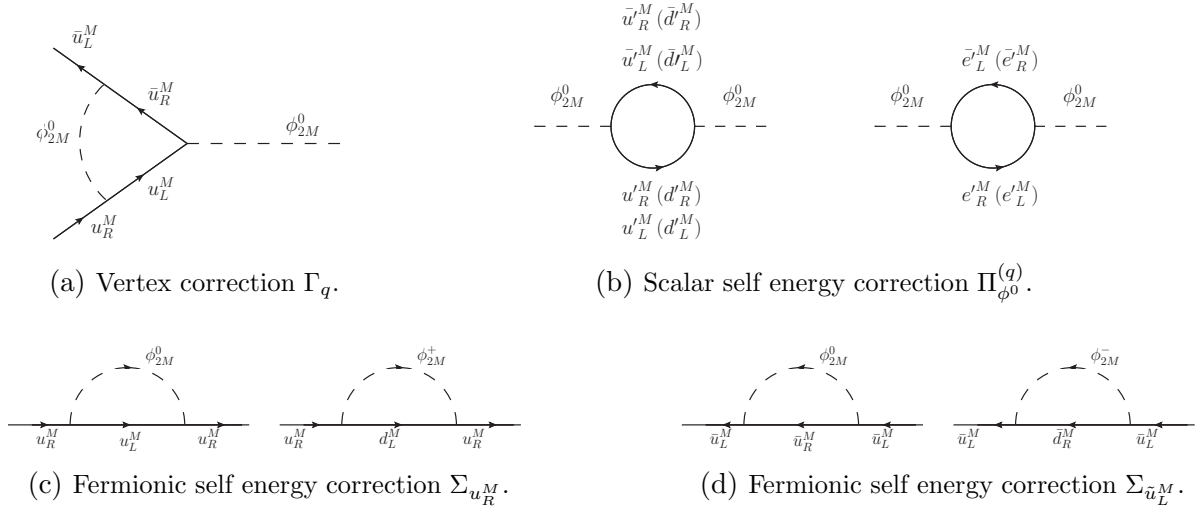
\begin{figure}[ht]
     \centering
     \begin{subfigure}[b]{0.3\textwidth}
         \centering
\scalebox{.45}{ 
\fcolorbox{white}{white}{
  \begin{picture}(244,182) (143,-123)
    \SetWidth{1.0}
    \SetColor{Black}
    \Line[arrow,arrowpos=0.65,arrowlength=5,arrowwidth=2,arrowinset=0.2](144,-122)(256,-42)
    \Line[arrow,arrowpos=0.20,arrowlength=5,arrowwidth=2,arrowinset=0.2](144,-122)(256,-42)
    \Line[arrow,arrowpos=0.80,arrowlength=5,arrowwidth=2,arrowinset=0.2](256,-42)(144,38)
    \Line[arrow,arrowpos=0.35,arrowlength=5,arrowwidth=2,arrowinset=0.2](256,-42)(144,38)
    \Line[dash,dashsize=10](256,-42)(384,-42)
    \Arc[dash,dashsize=10](232,-40)(66,135,228)
    \Text(160,38)[lb]{\Large{\Black{$\bar{u}_L^M $}}}
    \Text(224,-10)[lb]{\Large{\Black{$\bar{u}_R^M $}}}
    \Text(340,-36)[lb]{\Large{\Black{$\phi_{2M}^{0}$}}}
    \Text(130,-42)[lb]{\Large{\Black{$\phi_{2M}^{0}$}}}
    \Text(224,-90)[lb]{\Large{\Black{$u_L^M $}}}
    \Text(176,-122)[lb]{\Large{\Black{${u}_R^M $}}}
  \end{picture}
}}
         \caption{\footnotesize Vertex correction $\Gamma_q$.}
         \label{fig:Feynman-Diagrams-vertex-quarks}
     \end{subfigure}
     \begin{subfigure}[b]{0.6\textwidth}
         \centering
\scalebox{.45}{ 
\fcolorbox{white}{white}{
  \begin{picture}(335,166) (143,-153)
    \SetWidth{1.0}
    \SetColor{Black}
    \Text(375,-42)[lb]{\Large{\Black{$\phi_{2M}^{0}$}}}
    \Line[dash,dashsize=10](216,-58)(280,-58)
    \Arc[arrow,arrowpos=0.45,arrowlength=5,arrowwidth=2,arrowinset=0.2](315,-58)(35,117,477)
    \Arc[arrow,arrowpos=0.92,arrowlength=5,arrowwidth=2,arrowinset=0.2](315,-58)(35,117,477)
    \Line[dash,dashsize=10](350,-58)(410,-58)
    \Text(230,-42)[lb]{\Large{\Black{$\phi_{2M}^{0}$}}}
    \Text(290,+23)[lb]{\Large{\Black{$\bar{u^\prime}_R^M \, (\bar{d^\prime}_R^M)$}}}
    \Text(290,-12)[lb]{\Large{\Black{$\bar{u^\prime}_L^M \, (\bar{d\prime}_L^M)$}}}
    \Text(290,-125)[lb]{\Large{\Black{${u^\prime}_R^M \, ({d^\prime}_R^M)$}}}
    \Text(290,-153)[lb]{\Large{\Black{${u^\prime}_L^M \, ({d^\prime}_L^M)$}}}
    \end{picture}  \hspace{-2.0cm}
  
    \begin{picture}(335,166) (143,-153)
    \SetWidth{1.0}
    \SetColor{Black}
    \Text(375,-42)[lb]{\Large{\Black{$\phi_{2M}^{0}$}}}
    \Line[dash,dashsize=10](216,-58)(280,-58)
    \Arc[arrow,arrowpos=0.45,arrowlength=5,arrowwidth=2,arrowinset=0.2](315,-58)(35,117,477)
    \Arc[arrow,arrowpos=0.92,arrowlength=5,arrowwidth=2,arrowinset=0.2](315,-58)(35,117,477)
    \Line[dash,dashsize=10](350,-58)(410,-58)
    \Text(230,-42)[lb]{\Large{\Black{$\phi_{2M}^{0}$}}}
    \Text(290,-12)[lb]{\Large{\Black{$\bar{e^\prime}_L^{M} \, (\bar{e^\prime}_R^M)$}}}
    \Text(290,-125)[lb]{\Large{\Black{${e^\prime}_R^M \, ({e^\prime}_L^M) $}}}
  \end{picture}
}}
         \caption{\footnotesize Scalar self energy correction $\Pi_{\phi^{0}}^{(q)}$.}
         \label{fig:Feynman-Diagrams-scalar-self-energy-quarks}
     \end{subfigure}
     \begin{subfigure}[b]{0.49\textwidth}
     \vspace{0.5cm}
         \centering
\scalebox{.45}{ 
\fcolorbox{white}{white}{
   \begin{picture}(192,16) (115,-31)
    \SetWidth{1.0}
    \SetColor{Black}
    \Line[arrow,arrowpos=0.15,arrowlength=5,arrowwidth=2,arrowinset=0.2](108,-80)(310,-80)
    \Line[arrow,arrowpos=0.45,arrowlength=5,arrowwidth=2,arrowinset=0.2](128,-80)(304,-80)
    \Line[arrow,arrowpos=0.90,arrowlength=5,arrowwidth=2,arrowinset=0.2](128,-80)(304,-80)
    \Arc[dash,dashsize=10,arrow,arrowpos=0.5,arrowlength=5,arrowwidth=2,arrowinset=0.2,clock](208,-80)(48,-180,-360)
    \Text(224,-30)[lb]{\large{\Black{$\phi_{2M}^{0}$}}}
    \Text(132,-100)[lb]{\large{\Black{$u_R^M$}}}
    \Text(202,-102)[lb]{\large{\Black{${u}_L^M$}}}
    \Text(268,-100)[lb]{\large{\Black{$u_R^M$}}}
  \end{picture}
 \hspace{1.0cm}
  \begin{picture}(192,16) (115,-31)
    \SetWidth{1.0}
    \SetColor{Black}
    \Line[arrow,arrowpos=0.15,arrowlength=5,arrowwidth=2,arrowinset=0.2](108,-80)(310,-80)
    \Line[arrow,arrowpos=0.45,arrowlength=5,arrowwidth=2,arrowinset=0.2](128,-80)(304,-80)
    \Line[arrow,arrowpos=0.90,arrowlength=5,arrowwidth=2,arrowinset=0.2](128,-80)(304,-80)
    \Arc[dash,dashsize=10,arrow,arrowpos=0.5,arrowlength=5,arrowwidth=2,arrowinset=0.2,,clock](208,-80)(48,-180,-360)
    \Text(224,-30)[lb]{\large{\Black{$\phi_{2M}^+$}}}
    \Text(112,-100)[lb]{\large{\Black{$ {u}_R^M$}}}
    \Text(192,-100)[lb]{\large{\Black{$ {d}_L^M$}}}
    \Text(262,-100)[lb]{\large{\Black{$ {u}_R^M$}}}
  \end{picture} 
}} \vspace{1.0cm}
         \caption{\footnotesize Fermionic self energy correction $\Sigma_{u_R^M} $.}
         \label{fig:Feynman-Diagrams-fermionic-self-energy-quarks-right}
     \end{subfigure}
     \begin{subfigure}[b]{0.49\textwidth}
     \vspace{0.5cm}
         \centering
\scalebox{.45}{ 
\fcolorbox{white}{white}{
  \begin{picture}(192,16) (115,-31)
    \SetWidth{1.0}
    \SetColor{Black}
    \Line[arrow,arrowpos=0.15,arrowlength=5,arrowwidth=2,arrowinset=0.2](310,-80)(108,-80)
    \Line[arrow,arrowpos=0.55,arrowlength=5,arrowwidth=2,arrowinset=0.2](304,-80)(128,-80)
    \Line[arrow,arrowpos=0.90,arrowlength=5,arrowwidth=2,arrowinset=0.2](304,-80)(128,-80)
    \Arc[dash,dashsize=10,arrow,arrowpos=0.5,arrowlength=5,arrowwidth=2,arrowinset=0.2](208,-80)(48,-360,-180)
    \Text(224,-30)[lb]{\large{\Black{$\phi_{2M}^{0}$}}}
    \Text(132,-100)[lb]{\large{\Black{$\bar{u}_L^M$}}}
    \Text(202,-102)[lb]{\large{\Black{$\bar{u}_R^M$}}}
    \Text(268,-100)[lb]{\large{\Black{$\bar{u}_L^M$}}}
  \end{picture} \hspace{1.0cm}
  \begin{picture}(192,16) (115,-31)
    \SetWidth{1.0}
    \SetColor{Black}
    \Line[arrow,arrowpos=0.15,arrowlength=5,arrowwidth=2,arrowinset=0.2](310,-80)(108,-80)
    \Line[arrow,arrowpos=0.55,arrowlength=5,arrowwidth=2,arrowinset=0.2](304,-80)(128,-80)
    \Line[arrow,arrowpos=0.90,arrowlength=5,arrowwidth=2,arrowinset=0.2](304,-80)(128,-80)
    \Arc[dash,dashsize=10,arrow,arrowpos=0.5,arrowlength=5,arrowwidth=2,arrowinset=0.2,clock,flip](208,-80)(48,-180,-360)
    \Text(224,-30)[lb]{\large{\Black{$\phi_{2M}^-$}}}
    \Text(112,-100)[lb]{\large{\Black{$ \bar{u}_L^M$}}}
    \Text(192,-100)[lb]{\large{\Black{$ \bar{d}_R^M$}}}
    \Text(262,-100)[lb]{\large{\Black{$ \bar{u}_L^M$}}}
  \end{picture}
}} \vspace{1.0cm}
         \caption{\footnotesize Fermionic self energy correction $\Sigma_{\tilde{u}_L^M}$.}
         \label{fig:Feynman-Diagrams-fermionic-self-energy-quark-left}
     \end{subfigure}
             \caption{\small One loop radiative corrections to the process {\footnotesize $u_{R}^{M}(p) + \overline{u}_{L}^{M}(p^\prime) \to \phi_{2M}^{0}(q)$}, where $u$ stands for a specific mirror up quark flavor, $u^\prime$, $d^\prime$ and $e^\prime$ stand each for respectively the three mirror up quark, three mirror down quark, and three mirror lepton flavors}.
        \label{fig:Feynman-diagrams-quark-corrections}
\end{figure}
\begin{figure}[ht]
     \centering
     \begin{subfigure}[b]{0.45\textwidth}
         \centering
\scalebox{.50}{ 
\fcolorbox{white}{white}{

  \begin{picture}(244,182) (143,-123)
    \SetWidth{1.0}
    \SetColor{Black}
    \Line[arrow,arrowpos=0.65,arrowlength=5,arrowwidth=2,arrowinset=0.2](144,-122)(256,-42)
    \Line[arrow,arrowpos=0.20,arrowlength=5,arrowwidth=2,arrowinset=0.2](144,-122)(256,-42)
    \Line[arrow,arrowpos=0.80,arrowlength=5,arrowwidth=2,arrowinset=0.2](256,-42)(144,38)
    \Line[arrow,arrowpos=0.35,arrowlength=5,arrowwidth=2,arrowinset=0.2](256,-42)(144,38)
    \Line[dash,dashsize=10](256,-42)(384,-42)
    \Arc[dash,dashsize=10](232,-40)(66,135,228)
	\Text(224,-90)[lb]{\Large{\Black{$\nu_{e_R}^{M,T}$}}}  
	\Text(176,-122)[lb]{\Large{\Black{$\nu_{e_R}^{M}$}}}
	\Text(224,-10)[lb]{\Large{\Black{$\nu_{e_R}^{M}$ }}}
	\Text(170,28)[lb]{\Large{\Black{$\nu_{e_R}^{M,T}$}}}
    \Text(322,-36)[lb]{\Large{\Black{$\chi^0$}}}
    \Text(144,-50)[lb]{\Large{\Black{$\chi^0$}}}
  \end{picture} 
  
}}
         \caption{\footnotesize Vertex correction $\Gamma_{\nu}$.}
         \label{fig:Feynman-Diagrams-vertex-neutrinos}
     \end{subfigure}
     \begin{subfigure}[b]{0.45\textwidth}
         \centering
\scalebox{.50}{ 
\fcolorbox{white}{white}{

  \begin{picture}(308,166) (143,-139)
    \SetWidth{1.0}
    \SetColor{Black}
    \Text(386,-50)[lb]{\Large{\Black{$\chi^0$}}}
    \Line[dash,dashsize=10](210,-58)(283,-58)
    \Arc[arrow,arrowpos=0.45,arrowlength=5,arrowwidth=2,arrowinset=0.2](320,-58)(35,117,477)
    \Arc[arrow,arrowpos=0.92,arrowlength=5,arrowwidth=2,arrowinset=0.2](320,-58)(35,117,477)
    \Line[dash,dashsize=10](357,-58)(418,-58)
    \Text(242,-50)[lb]{\Large{\Black{$\chi^0$}}}
    \Text(315,-12)[lb]{\Large{\Black{$\nu_{e_R^\prime}^M$}}}
    \Text(315,-125)[lb]{\Large{\Black{$\nu_{e_R^\prime}^{M\,T}$}}}
  \end{picture}
  
}}
         \caption{\footnotesize Scalar self energy correction $\Pi_{\chi^{0} }^{(\nu_{_R})}$.}
         \label{fig:Feynman-Diagrams-scalar-self-energy-neutrinos}
     \end{subfigure}
     \begin{subfigure}[b]{0.90\textwidth}
         \centering
\scalebox{.60}{ 
\fcolorbox{white}{white}{

  \begin{picture}(192,96) (115,-105)
    \SetWidth{1.0}
    \SetColor{Black}
    \Line[arrow,arrowpos=0.15,arrowlength=5,arrowwidth=2,arrowinset=0.2](108,-80)(310,-80)
    \Line[arrow,arrowpos=0.45,arrowlength=5,arrowwidth=2,arrowinset=0.2](128,-80)(304,-80)
    \Line[arrow,arrowpos=0.90,arrowlength=5,arrowwidth=2,arrowinset=0.2](128,-80)(304,-80)
    \Arc[dash,dashsize=10,arrow,arrowpos=0.5,arrowlength=5,arrowwidth=2,arrowinset=0.2,clock](208,-80)(48,-180,-360)
    \Text(204,-25)[lb]{\large{\Black{$\chi^0$}}}
    \Text(120,-105)[lb]{\large{\Black{$\nu_{e_R}^M $}}}
    \Text(197,-105)[lb]{\large{\Black{$\nu_{e_R}^{M\,T} $}}}
    \Text(282,-105)[lb]{\large{\Black{$\nu_{e_R}^M$}}}
  \end{picture} \hspace{1.0cm}
  
  \begin{picture}(192,96) (115,-105)
    \SetWidth{1.0}
    \SetColor{Black}
    \Line[arrow,arrowpos=0.15,arrowlength=5,arrowwidth=2,arrowinset=0.2](108,-80)(310,-80)
    \Line[arrow,arrowpos=0.45,arrowlength=5,arrowwidth=2,arrowinset=0.2](128,-80)(304,-80)
    \Line[arrow,arrowpos=0.90,arrowlength=5,arrowwidth=2,arrowinset=0.2](128,-80)(304,-80)
    \Arc[dash,dashsize=10,arrow,arrowpos=0.5,arrowlength=5,arrowwidth=2,arrowinset=0.2,clock](208,-80)(48,-180,-360)
    \Text(204,-25)[lb]{\large{\Black{$\chi^+$}}}
    \Text(120,-105)[lb]{\large{\Black{$ \nu_{e_R}^M$}}}
    \Text(195,-105)[lb]{\large{\Black{$e_R^{M\,T}$}}}
    \Text(270,-105)[lb]{\large{\Black{$\nu_{e_R}^M$}}}
  \end{picture} \hspace{1.0cm}
  
  \begin{picture}(192,96) (115,-105)
    \SetWidth{1.0}
    \SetColor{Black}
    \Line[arrow,arrowpos=0.15,arrowlength=5,arrowwidth=2,arrowinset=0.2](108,-80)(310,-80)
    \Line[arrow,arrowpos=0.45,arrowlength=5,arrowwidth=2,arrowinset=0.2](128,-80)(304,-80)
    \Line[arrow,arrowpos=0.90,arrowlength=5,arrowwidth=2,arrowinset=0.2](128,-80)(304,-80)
    \Arc[dash,dashsize=10,arrow,arrowpos=0.5,arrowlength=5,arrowwidth=2,arrowinset=0.2,clock](208,-80)(48,-180,-360)
    \Text(204,-25)[lb]{\large{\Black{$\phi_{2M}^+$}}}
    \Text(132,-105)[lb]{\large{\Black{$\nu_{e_R}^M$}}}
    \Text(205,-105)[lb]{\large{\Black{$e_L^M$}}}
    \Text(278,-105)[lb]{\large{\Black{$\nu_{e_R}^M$}}}
  \end{picture}
}}
         \caption{\footnotesize Fermionic self energy correction $\Sigma_{\nu_{_R}^M}$.}
         \label{fig:Feynman-Diagrams-fermionic-self-energy-neutrinos}
     \end{subfigure}

     \begin{subfigure}[b]{0.90\textwidth}
         \centering

\scalebox{.60}{ 
\fcolorbox{white}{white}{

  \begin{picture}(192,96) (115,-105)
    \SetWidth{1.0}
    \SetColor{Black}
    \Line[arrow,arrowpos=0.15,arrowlength=5,arrowwidth=2,arrowinset=0.2](108,-80)(310,-80)
    \Line[arrow,arrowpos=0.45,arrowlength=5,arrowwidth=2,arrowinset=0.2](128,-80)(304,-80)
    \Line[arrow,arrowpos=0.90,arrowlength=5,arrowwidth=2,arrowinset=0.2](128,-80)(304,-80)
    \Arc[dash,dashsize=10,arrow,arrowpos=0.5,arrowlength=5,arrowwidth=2,arrowinset=0.2,clock](208,-80)(48,-180,-360)
    \Text(204,-25)[lb]{\large{\Black{$\chi^0$}}}
    \Text(120,-105)[lb]{\large{\Black{$\nu_{e_R}^{M,T}$}}}
    \Text(197,-105)[lb]{\large{\Black{$ \nu_{e_R}^M$}}}
    \Text(272,-105)[lb]{\large{\Black{$\nu_{e_R}^{M,T}$}}}
  \end{picture} \hspace{1.0cm}
  
  \begin{picture}(192,96) (115,-105)
    \SetWidth{1.0}
    \SetColor{Black}
    \Line[arrow,arrowpos=0.15,arrowlength=5,arrowwidth=2,arrowinset=0.2](108,-80)(310,-80)
    \Line[arrow,arrowpos=0.45,arrowlength=5,arrowwidth=2,arrowinset=0.2](128,-80)(304,-80)
    \Line[arrow,arrowpos=0.90,arrowlength=5,arrowwidth=2,arrowinset=0.2](128,-80)(304,-80)
    \Arc[dash,dashsize=10,arrow,arrowpos=0.5,arrowlength=5,arrowwidth=2,arrowinset=0.2,clock](208,-80)(48,-180,-360)
    \Text(204,-25)[lb]{\large{\Black{$\chi^+$}}}
    \Text(120,-105)[lb]{\large{\Black{$\nu_{e_R}^{M,T}$}}}
    \Text(190,-105)[lb]{\large{\Black{$e_R^M $}}}
    \Text(270,-105)[lb]{\large{\Black{$\nu_{e_R}^{M,T}$}}}
  \end{picture}
}}
         \caption{\footnotesize Fermionic self energy correction $\Sigma_{\nu_{_R}^{M,T}}$.}
         \label{fig:Feynman-Diagrams-fermionic-self-energy-neutrinos-T}
     \end{subfigure}
             \caption{\small One loop radiative corrections to the process {\footnotesize $ \nu_{e_R}(p) + \nu_{e_R}^{T}(p^\prime) \to \chi^{0}(q)$}, where $e$ stands for a specific mirror lepton flavor, and $e^\prime$ stands for the three mirror lepton flavors.}
        \label{fig:Feynman-diagrams-neutrino-corrections}
\end{figure}
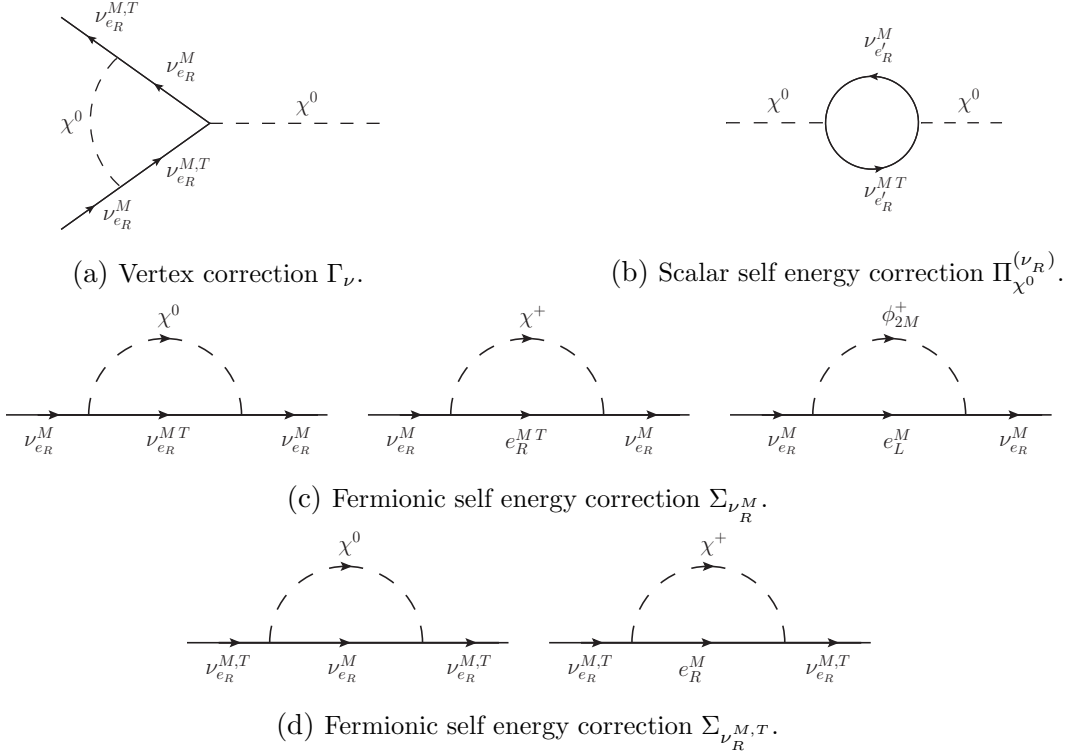
\begin{itemize}
    \item The $\Gamma _{e}$, $\Gamma_{q}$ and $\Gamma_{\nu}$ vertex one loop corrections: 
\begin{equation}
    \Gamma _{e}(p, p^\prime, \theta, a, b, g_{e^{M}}, \tilde{g}_{e^{M}} )  = I \left( p, p^\prime, \theta, a, b, g_{e^{M}}, \tilde{g}_{e^{M}} \right),
\end{equation}
\begin{equation}
    \Gamma _{q}(p, p^\prime, \theta, a, b, g_{q^{M}}, \tilde{g}_{q^{M}} )  = I \left( p, p^\prime, \theta, a, b, g_{q^{M}}, \tilde{g}_{q^{M}} \right),
\end{equation}
\begin{equation}
    \Gamma _{\nu}(p, p^\prime, \theta, a, b,  g_{_{M}}, \tilde{g}_{_{M}} )  = I \left( p, p^\prime, \theta, a, b,  g_{_{M}}, \tilde{g}_{_{M}} \right),
\end{equation}
where:
\begin{eqnarray}
I \left( p, p^\prime, \theta, a, b, g_{_X}, \tilde{g}_{_X} \right) &=&\int \frac{d^{\scriptscriptstyle D}k}{\left( 2\pi \right)^{\scriptscriptstyle D}} \, \,(-1)
\left( g_{_X} e^{-\frac{i}{2}(p-p^\prime)\tilde{k}}+ \tilde{g}_{_X} e^{+\frac{i}{2}(p-p^\prime)\tilde{k}}\right)
\left( g_{_X} e^{\frac{i}{2}p^\prime(\tilde{k}-\tilde{p})} \right. 
 \nonumber \\
&& \left. + \,\tilde{g}_{_X} e^{-\frac{i}{2}p^\prime(\tilde{k}-\tilde{p})}\right) \times \left( g_{_X} e^{\frac{i}{2}k\tilde{p}}+ \tilde{g}_{_X} e^{-\frac{i}{2}k\tilde{p}}\right)
\times \frac{ i\mathbf{\kslash}
+ib\frac{\tkslash}{\theta^2 k^{2}} }
{ k^{2}+\frac{b^{2}}{\theta^2 k^{2}} }
 \nonumber \\
&& \times 
\frac{ i\left( \mathbf{\kslash}-\pslash +\ppslash \right) +ib\frac{  \tkslash -\tpslash + \tppslash }{\theta^2\left( k-p+p^\prime\right) ^{2}}}
{ \left(k-p + p^\prime\right)^{2}+\frac{b^{2}}{\theta^2  \left( k-p+p^\prime\right) ^{2}}} \times 
\frac{1}{ \left( k-p\right) ^{2}+\frac{a^{2}}{\theta^2 \left( k-p\right) ^{2}} }\,.
\end{eqnarray}
\item The $\Sigma_{e}$ mirror down leptonic self energy one loop correction : 
\begin{eqnarray}
\centering
\Sigma_{e}\left( p, p^\prime, \theta, a, b, g_{e^{M}}, \tilde{g}_{e^{M}}, g_{_M}, \tilde{g}_{_M} \right) &=&  \Sigma_{{e}_R^M} \left( p, \theta, a, b, g_{e^{M}}, \tilde{g}_{e^{M}}, g_{_M}, \tilde{g}_{_M} \right) 
  \nonumber \\ && \qquad + \,\,\, \Sigma_{\tilde{e}_L^M}\left(p^\prime, \theta, a, b, g_{e^{M}}, \tilde{g}_{e^{M}} \right), 
\end{eqnarray}
with:
\begin{equation} 
    \Sigma_{{e}_R^M} \left( p, \theta, a, b, g_{e^{M}}, \tilde{g}_{e^{M}}, g_{_M}, \tilde{g}_{_M} \right) =  J \left( p, \theta, a, b, g_{e^{M}}, \tilde{g}_{e^{M}} \right) 
    +  \frac{3}{2}J \left( p, \theta, a, b, g_{_M}, \tilde{g}_{_M} \right),
\end{equation}
and 
\begin{equation}
    \Sigma_{\tilde{e}_L^M}\left(p^\prime, \theta, a, b, g_{e^{M}}, \tilde{g}_{e^{M}} \right) = 2\, J \left(p^\prime, \theta, a, b, g_{e^{M}}, \tilde{g}_{e^{M}} \right),
\end{equation}

where:
\begin{eqnarray}
    J \left( l, \theta, a, b, g_{_X}, \tilde{g}_{_X} \right) &=& \int \frac{d^{\scriptscriptstyle D}k}{\left( 2\pi \right) ^{\scriptscriptstyle D}} \,  \left[ g_{_X}e^{-\frac{i}{2} l \tilde{k}}+\tilde{g}_{_X}e^{+\frac{i}{2} l \tilde{k}}\right] \left[ g_{_X}e^{-\frac{i}{2}k\tilde{l}}+\tilde{g}_{_X}e^{+\frac{i}{2}k\tilde{l}}\right]  \nonumber \\
&& \qquad\qquad \times \left( \frac{i\left( \lslash + \kslash \right) + ib\frac{ \tkslash + \tlslash }{\theta^2 \left(l+k\right)^{2}}}{\left( l+k\right) ^{2}+\frac{b^{2}}{\theta^2 \left(l+k\right) ^{2}}}\right) \frac{1}{k^{2}+\frac{a^{2}}{\theta^2 k^{2}}}\,.
\end{eqnarray}
\item The  $\Sigma_{q}$ mirror quark self energy one loop correction :
\begin{equation}
   \Sigma_{q}\left( p, p^\prime, \theta, a, b, g_{q^{M}}, \tilde{g}_{q^{M}} \right) =  \Sigma_{{q}_R^M} \left( p, \theta, a, b , g_{q^{M}}, \tilde{g}_{q^{M}}\right) + \Sigma_{\tilde{e}_L^M}\left(p^\prime, \theta, a, b , g_{q^{M}}, \tilde{g}_{q^{M}} \right),
\end{equation}
with
\begin{equation}
    \Sigma_{{q}_R^M} \left( p, \theta, a, b, g_{q^{M}}, \tilde{g}_{q^{M}} \right) = 2\, J \left(p, \theta, a, b, g_{q^{M}}, \tilde{g}_{q^{M}} \right),
\end{equation}
and 
\begin{equation}
    \Sigma_{\tilde{e}_L^M}\left(p^\prime, \theta, a, b, g_{q^{M}}, \tilde{g}_{q^{M}} \right) = 2\, J \left(p^\prime, \theta, a, b, g_{q^{M}}, \tilde{g}_{q^{M}} \right).
\end{equation}
\item The $\Sigma_{\nu}$ right handed neutrinos' self energy one loop corrections : 
\begin{eqnarray}
\Sigma_{\nu}\left( p, p^\prime, \theta, a, b, g_{_M}, \tilde{g}_{_M}, g_{e^{M}}, \tilde{g}_{e^{M}} \right) 
&=&  \Sigma_{{\nu}_R^M}\left( p, \theta, a, b, g_{_M}, \tilde{g}_{_M}, g_{e^{M}}, \tilde{g}_{e^{M}} \right) 
\nonumber \\
&& \qquad +\,\,\, \Sigma_{{\nu}_R^{M,T}}\left(p^\prime, \theta, a, b, g_{_M}, \tilde{g}_{_M} \right),
\end{eqnarray}
with:
\begin{equation} 
    \Sigma_{{\nu}_R^M} \left( p, \theta, a, b, g_{_M}, \tilde{g}_{_M}, g_{e^{M}}, \tilde{g}_{e^{M}} \right) =  \frac{3}{2}\,J \left( p, \theta, a, b, g_{_M}, \tilde{g}_{_M} \right) + J \left( p, \theta, a, b, g_{e^{M}}, \tilde{g}_{e^{M}} \right),
\end{equation}
and 
\begin{equation}
    \Sigma_{{\nu}_R^{M,T}}\left(p^\prime, \theta, a, b, g_{_M}, \tilde{g}_{_M} \right) =  \frac{3}{2}\,J \left(p^\prime, \theta, a, b, g_{_M}, \tilde{g}_{_M} \right).
\end{equation}

\item And the $\Pi_{\phi^{0}}^{(e)}$, $\Pi_{\phi^{0}}^{(q)}$ and $\Pi_{\chi^{0} }^{(\nu_{_R})}$ scalar self energy one loop correction :
\begin{eqnarray}
    &&    \Pi_{\phi^{0}}^{(e)} \left( p-p^\prime, \theta, b, g_{e^{M}}, \tilde{g}_{e^{M}}, g_{q^{M}}, \tilde{g}_{q^{M}} \right)=\Pi_{\phi^{0}}^{(q)} \left( p-p^\prime, \theta, b, g_{e^{M}}, \tilde{g}_{e^{M}}, g_{q^{M}}, \tilde{g}_{q^{M}}\right) 
    \nonumber \\
    && \qquad \qquad  = 6\, K \left(p-p^\prime, \theta, b, g_{e^{M}}, \tilde{g}_{e^{M}} \right) + 12 \, K \left(p-p^\prime, \theta, b, g_{q^{M}}, \tilde{g}_{q^{M}} \right),
\end{eqnarray}
\begin{equation}
  \Pi_{\chi^{0} }^{(\nu_{_R})} \left( p-p^\prime,  \theta, b, g_{_{M}}, \tilde{g}_{_{M}} \right) = 3\,  K \left(p-p^\prime, \theta, b, g_{_{M}}, \tilde{g}_{_{M}} \right),
\end{equation}
where
\begin{eqnarray}
K \left(l, \theta, b, g_{_X}, \tilde{g}_{_X} \right) &=& \int \dfrac{d^{\scriptscriptstyle D}k}{\left( 2\pi \right)^{\scriptscriptstyle D}} \left( -1\right)  \left[ g_{_X}e^{-\frac{i}{2} l\tilde{k}}+\tilde{g}_{_X}e^{+\frac{i}{2} l \tilde{k}}\right] \left[ g_{_X}e^{-\frac{i}{2}k\tilde{l}}+\tilde{g}_{_X}e^{+\frac{i}{2} k \tilde{l}} \right] \nonumber \\
&& \times \, tr\left[ \left( \frac{i \kslash +ib\frac{\tkslash}{\theta^2 k^{2}}}{k^{2}+\frac{b^{2}}{\theta^2 k^{2}}}\right)  \left( \dfrac{i \left( \lslash + \kslash \right) + i b \frac{ \tkslash + \tlslash }{\theta^2 \left( l+k\right) ^{2}}}{\left( l+k\right) ^{2}+ \frac{b^{2}}{ \theta^2 \left( l+k\right) ^{2}}}\right) \right].
\end{eqnarray}
\end{itemize}
\noindent Each of these D dimension Euclidean momentum space integrals can be sundered into planar and non-planar integrals. Hence, performing this calculation using the dimensional regularization method and in the way described in \cite{Bouchachia-2015, BGKRSS-2008} leads us, for our special case of massless particles, to the following UV divergent part:
\begin{eqnarray}
\Gamma_{e}(p, p^\prime, \theta, a, b, g_{e^{M}}, \tilde{g}_{e^{M}}) &=& -\dfrac{2 \,g_{e^{M}} \, \tilde{g}_{e^{M}}}{\left( 4\pi \right)^{2}\varepsilon_{_{UV}} } \left\{ g_{e^{M}} \, e^{-\frac{i}{2} p^\prime \tilde{p}} + \tilde{g}_{e^{M}} \, e^{+\frac{i}{2} p^\prime \tilde{p}} \right\} \nonumber \\
&& \qquad + \, \text{UV finite part}(p, p^\prime, \theta, a, b, g_{e^{M}}, \tilde{g}_{e^{M}}),
\end{eqnarray}
\begin{eqnarray}
\Gamma_{q}(p, p^\prime, \theta, a, b, g_{q^{M}}, \tilde{g}_{q^{M}})&=& 
-\dfrac{2 \,g_{q^{M}} \, \tilde{g}_{q^{M}}}{\left( 4\pi \right)^{2}\varepsilon_{_{UV}} } \left\{ g_{q^{M}} \, e^{-\frac{i}{2}(p^\prime \tilde{p}}  + \tilde{g}_{q^{M}} \, e^{+\frac{i}{2} p^\prime \tilde{p}} \right\} \nonumber \\
&& \qquad + \, \text{UV finite part}(p, p^\prime, \theta, a, b, g_{q^{M}}, \tilde{g}_{q^{M}}),
\end{eqnarray}
\begin{eqnarray}
\Gamma_{\nu}(p, p^\prime, \theta, a, b, g_{_{M}}, \tilde{g}_{_{M}})&=& -\dfrac{2\,g_{_M} \, \tilde{g}_{_M}}{\left( 4\pi \right) ^{2}\varepsilon_{_{UV}} } \left\{ g_{_M} \, e^{-\frac{i}{2} p^\prime \tilde{p}} + \tilde{g}_{_M} \, e^{+\frac{i}{2} p^\prime \tilde{p}} \right\} \nonumber \\ 
&& \qquad + \, \text{UV finite part}(p, p^\prime, \theta, a, b, g_{_{M}}, \tilde{g}_{_{M}}),
\end{eqnarray}
\begin{eqnarray}
  &&  \Sigma_{e}(p, p^\prime, \theta, a, b, g_{e^{M}}, \tilde{g}_{e^{M}}, g_{_M}, \tilde{g}_{_M}) =\frac{-i}{\left( 4\pi \right) ^{2}\varepsilon_{_{UV}} } \left\{\left[ \left( g_{e^{M}}^{2}+ \tilde{g}_{e^{M}}^{2}\right) + \frac{3}{2}\left( g_{_M}^{2}+\tilde{g}_{_M}^{2}\right) \right] \pslash \right. \nonumber \\
  && \qquad \bigg.  + 2\biggl[ g_{e^{M}}^{2}+ \tilde{g}_{e^{M}}^{2} \biggr] \ppslash \biggr\}
    + \,\,\, \text{UV finite part}(p, p^\prime, \theta, a, b, g_{e^{M}}, \tilde{g}_{e^{M}}, g_{_M}, \tilde{g}_{_M}),
\end{eqnarray}
\begin{equation}
\Sigma_{q}(p, p^\prime, \theta, a, b, g_{q^{M}}, \tilde{g}_{q^{M}}) =-\frac{ 2 i \left[ g_{q^{M}}^{2}+\tilde{g}_{q^{M}}^{2}\right] }{\left( 4\pi \right) ^{2}\varepsilon_{_{UV}} }\left( \pslash + \ppslash \right) + \, \text{UV finite part}(p, p^\prime, \theta, a, b, g_{q^{M}}, \tilde{g}_{q^{M}}),
\end{equation}
\begin{eqnarray}
 &&   \Sigma_{\nu} (p, p^\prime, \theta, a, b, g_{_M}, \tilde{g}_{_M}, g_{e^{M}}, \tilde{g}_{e^{M}}) = \frac{-i}{\left(4\pi \right)^{2}\varepsilon_{_{UV}} } \left\{\left[ \frac{3}{2}\left( g_{M}^{2}+\tilde{g}_{M}^{2}\right) +g_{e^{M}}^{2}+\tilde{g}_{e^{M}}^{2}\right]\pslash \right. \nonumber \\
 &&  \qquad + \left. \left[ \frac{3}{2}\left( g_{M}^{2}+\tilde{g}_{M}^{2}\right)\right]\ppslash \right\}   + \,\,\, \text{UV finite part}(p, p^\prime, \theta, a, b, g_{_M}, \tilde{g}_{_M}, g_{e^{M}}, \tilde{g}_{e^{M}}),
\end{eqnarray}
\begin{eqnarray}
  && \Pi_{\phi^0}^{(e)}(p-p^\prime, \theta, b, g_{e^{M}}, \tilde{g}_{e^{M}}, g_{q^{M}}, \tilde{g}_{q^{M}})=\Pi_{\phi^0}^{(q)}\left( p-p^\prime, \theta, b, g_{e^{M}}, \tilde{g}_{e^{M}}, g_{q^{M}}, \tilde{g}_{q^{M}} \right) \nonumber \\
&& \nonumber \\
&&\qquad \qquad \qquad = \, 3 \times
\frac{8\left(g_{e^{M}}^{2}+\tilde{g}_{e^{M}}^{2}\right) + 16\left(g_{q^{M}}^{2}+\tilde{g}_{q^{M}}^{2}\right)
}{\left( 4\pi \right)^{2} \varepsilon_{_{UV}}}   \, (p-p^\prime)^{2} \nonumber \\
&& \nonumber \\
&& \qquad \qquad \qquad \qquad \qquad + \,\,\, \text{UV finite part}(p-p^\prime, \theta, b, g_{e^{M}}, \tilde{g}_{e^{M}}, g_{q^{M}}, \tilde{g}_{q^{M}}),
\end{eqnarray}
\begin{equation}
\Pi_{\chi^0}^{(\nu)}(p-p^\prime,  \theta, b, g_{_{M}}, \tilde{g}_{_{M}}) = \, 3 \times \frac{4\left( g_{_M}^{2}+\tilde{g}_{_M}^{2}\right) }{ \left( 4\pi \right)^{2}\varepsilon_{_{UV}} } \, (p-p^\prime)^{2} + \, \text{UV finite part}(p-p^\prime,  \theta, b, g_{_{M}}, \tilde{g}_{_{M}}).
\end{equation}
We then introduce the divergent parts of these radiative corrections into Equations (\ref{conter-term-scalar-propagator}), (\ref{conter-term-fermionic-propagator}) and (\ref{conter-term-vertex}), to infer directly the corresponding counter terms given by Equations (\ref{equ-counter-term-gem-gemtilde}), (\ref{equ-counter-term-e}), (\ref{equ-counter-term-phi0e}), (\ref{equ-counter-term-gqm-gqmtilde}), (\ref{equ-counter-term-q}), (\ref{equ-counter-term-phi0q}), (\ref{equ-counter-term-gm-gmtilde}), (\ref{equ-counter-term-nuR}), and (\ref{equ-counter-term-chi0-nuR}).
\section*{Acknowledgement}
K. A. Bouteldja would like to acknowledge fruitful discussions with  V. Rivassau  and with M. Schweda at the beginning of this work. N. Bouayed would like to thank M. Mamou for helpful discussions about inverse problems and numerical solutions of differential equations.

\noindent Authors would like to thank P. Aurenche, B. Si Lakhal, and A. Yanallah for reading the manuscript and giving fruitful comments and suggestions, and A.M. Bouayed for proofreading the manuscript and revealing some grammatical errors and misspellings.

\let\doi\relax

\normalem
\bibliographystyle{natbib}

\end{document}